\documentclass[twoside,twocolumn,9pt]{article}

\usepackage{extsizes}
\usepackage[super,sort&compress,comma]{natbib} 
\usepackage[version=3]{mhchem}
\usepackage[left=1.5cm, right=1.5cm, top=1.785cm, bottom=2.0cm]{geometry}
\usepackage{balance}
\usepackage{mathptmx}
\usepackage{sectsty}
\usepackage{graphicx} 
\usepackage{lastpage}
\usepackage[format=plain,justification=justified,singlelinecheck=false,font={stretch=1.125,small,sf},labelfont=bf,labelsep=space]{caption}
\usepackage{float}
\usepackage{fancyhdr}
\usepackage{fnpos}
\usepackage[english]{babel}
\addto{\captionsenglish}{%
	
}
\usepackage{array}
\usepackage{droidsans}
\usepackage{charter}
\usepackage[T1]{fontenc}
\usepackage[usenames,dvipsnames]{xcolor}
\usepackage{setspace}
\usepackage[compact]{titlesec}
\usepackage{hyperref}

\let\OLDthebibliography\thebibliography
\renewcommand\thebibliography[1]{
	\OLDthebibliography{#1}
	\setlength{\parskip}{0pt}
	\setlength{\itemsep}{0pt plus 0.25ex}
}

\begin{document}

\title{Structure-property relationships and the mechanisms of multistep transitions in spin crossover materials and frameworks}
\author{Jace Cruddas$^*$ and B. J. Powell$^\dagger$}
\date{$^*$j.cruddas@uq.edu.au\\
	$^\dagger$powell@physics.uq.edu.au\\
	School of Mathematics and Physics$,$ The University of Queensland$,$ QLD 4072$,$ Australia}

\maketitle

\textit{\textbf{Abstract:} Spin crossover frameworks and molecular crystals display fascinating collective behaviours. This includes multi-step transitions with hysteresis and a wide variety of long-range ordered patterns of high-spin and low-spin metal centres. From both practical and fundamental perspectives it is important to understand the mechanisms behind these collective behaviours. We study a simple model of elastic interactions and identify thirty six different spin-state ordered phases. We observe spin-state transitions with between one and eight steps. These include both sharp transitions and crossovers, and both complete and incomplete spin crossover.  We demonstrate structure-property relationships that explain these differences. These arise because through-bond interactions are antiferroelastic (favour metal centres with different spin-states); whereas, through-space interactions are typically ferroelastic (favour the same spin-state). In general, rigid materials with longer range elastic interactions lead to transitions with more steps and more diverse spin-state ordering, which explains why both are prominent in frameworks.} 


\section*{Introduction}

Spin crossover (SCO) materials typically consist of a transition metal ion with a partially occupied d-shell 
surrounded by several ligands. Depending on the physical environment (temperature, pressure, magnetic field, exposure to light, \textit{etc.}) these materials can exist in either high spin (HS) or low spin (LS) states.\cite{gutlich} 

\begin{figure}[h]
	\centering
	\includegraphics[width=\columnwidth]{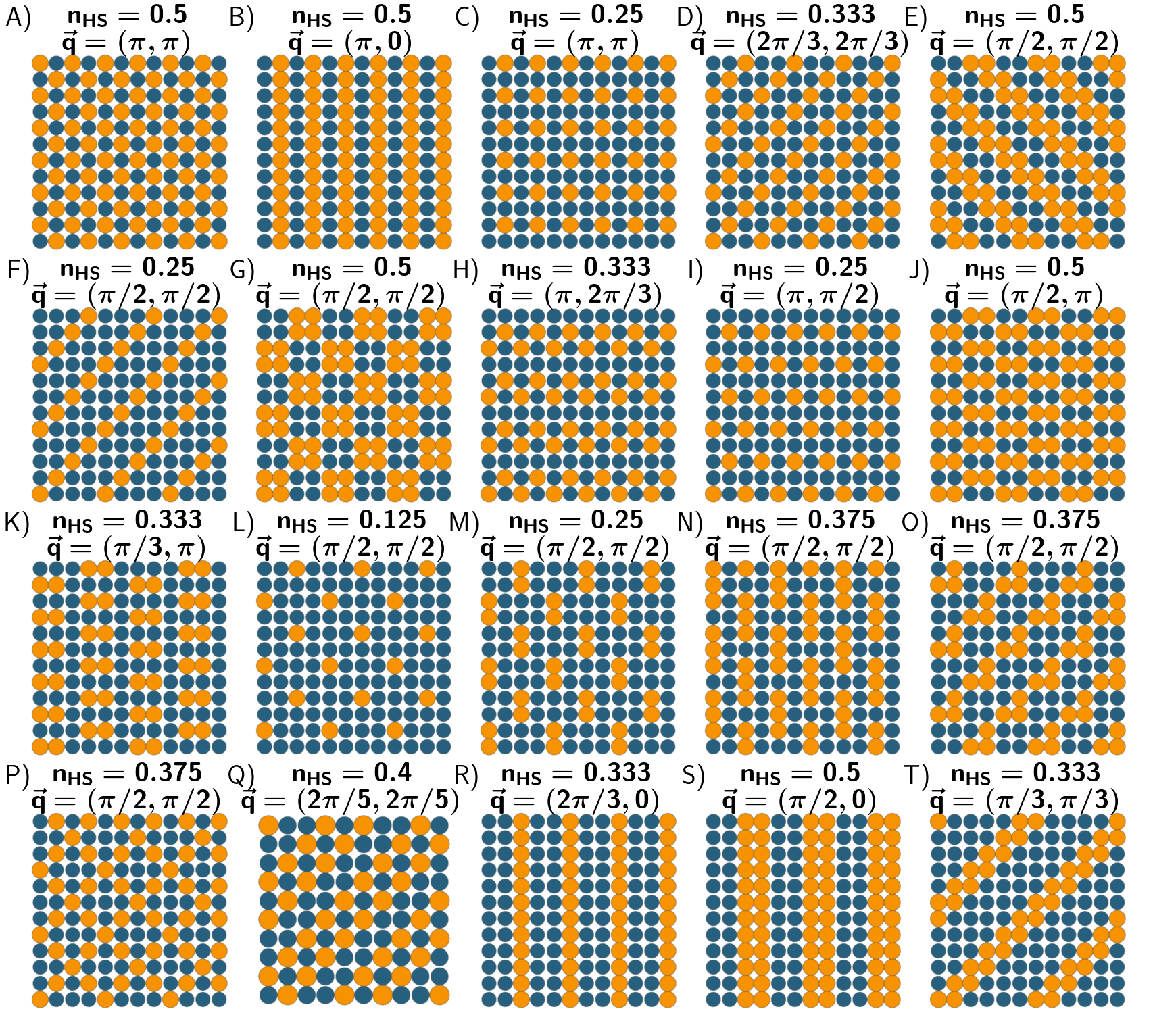} 
	\caption{ Truncated snapshots of the antiferroelastic spin-state orderings with majority or half LS metal centres (blue circles)  observed in our  Monte Carlo simulations on a 60$\times$60 lattice. Equivalent phases are found with majority HS (orange circles). For each state the fraction of HS ions, $n_{HS}$, and the Bragg wave vector, $\vec{q}$, are indicated. By analogy to magnetic order phase A is known as  N\'eel order and phase B is referred to as stripe order. 
	}
	\label{phases} 
\end{figure}

Metal-ligand bond lengths in the HS state are around $10~\%$ longer than in the LS state.\cite{gutlich}
In the solid state, long-range elastic interactions between metal centres couple to the local structural distortions caused by individual metal centres changing their spin-state. This causes a wide range of different thermodynamic behaviours, including first-order transitions with hysteresis,  incomplete transitions, crossovers, and up to eight-step transitions.\cite{Peng,Ortega,Ni,gutlich}
Many different long-range ordered patterns of HS and LS metal centres, collectively known as antiferroelastic phases (Fig. \ref{phases}; Table \ref{Table:Phases}), have been observed.\cite{Zenere,Milin,SciortinoIC,Liu17,Liu18,Murphy,Sciortino,Klein,SciortinoCS,lopez,Clements,Zhang,Meng,Agusti_2,Zhang_2,Liu2,Augusti3,Kosone,Halder,Lin,Adams,bao,Halder2,Hang,Wei,Luan,Breful,Matsumoto,Li,Money,Murnaghan,vieira,Chernyshov,klingele,nihei,Fitzpatrick,IsingWatanabe}

This leads one to ask what mechanism is responsible for these collective effects? and can one predict what other behaviours might exist? Beyond the fundamental interest in these questions, SCO materials and frameworks have been sought after for their many potential applications including high-density reversible memory \cite{gutlich,halcrow,KahnAM,Eric,Fouche,Bousseksou,Khan}, actuators \cite{Mikolaseq}, ultrafast nanoscale switches, \cite{gutlich,halcrow,Eric,Fouche,Bousseksou,Khan} sensors \cite{gutlich,Jureschi,Bousseksou,Khan} and displays \cite{gutlich,halcrow,KahnAM,Fouche,Bousseksou,Khan}. 
Understanding the  mechanisms that control the collective behaviours of SCO materials could significantly enhance their potential to be  engineered for specific applications.

Two-step transitions can arise as a result of molecular multistability, from crystallographic differences between metal centers or as a result of interactions between the spin-states of the metals. \cite{ColletReview} Interaction-driven two-step transtions have been studied extensively on the basis of empirical models. \cite{IsingBouk, IsingBous, IsingChern, IsingNishino, IsingWatanabe} 
Recently, it has been suggested that the competition between elastic interactions or `elastic frustration' is crucial for understanding two-step transitions and antiferroelastic order in SCO materials. \cite{Paez-Espejo, StretchBend, Jace} For example, a mismatch between the equilibrium bond lengths of nearest and next nearest-neighbour bonds, has been shown to induce two-step transitions on the square lattice. \cite{Paez-Espejo}
Similarly, it has been argued that in phenomenological Ising-like models cooperative nearest and next-nearest neighbour interactions are required stabilize two-step transitions and antiferroelastic order in SCO materials. \cite{IsingBouk, IsingBous, IsingChern, IsingNishino, IsingWatanabe} 

However, significantly less is known about the mechanisms of multi-step transitions.
The question of what mechanism is responsible for three-, four- and higher-step transitions, and the diversity of antiferroelastic orders reported remains open. In particular, the link between antiferroelastic order and the topology of the lattice remains largely unexplored.

Multi-step transitions have been reported for a variety of different molecular materials and frameworks on square arrays including Hofmann-type molecular frameworks \cite{Ni,Ortega,Zenere,Milin,SciortinoIC,Liu17,Liu18,Murphy,Sciortino,Klein,SciortinoCS,lopez,Clements,Zhang,Meng,Agusti_2,Zhang_2,Liu2,Augusti3,Kosone}, other coordination polymers\cite{Halder,Lin,Adams,bao,Halder2} and molecular crystals held together by supramolecular interactions\cite{Ortega,Hang,Wei,Luan,Breful,Matsumoto,Li,Money,Murnaghan,vieira,Chernyshov,klingele,nihei,Fitzpatrick}.

\begin{figure}
	\centering
	\includegraphics[width=0.85\columnwidth]{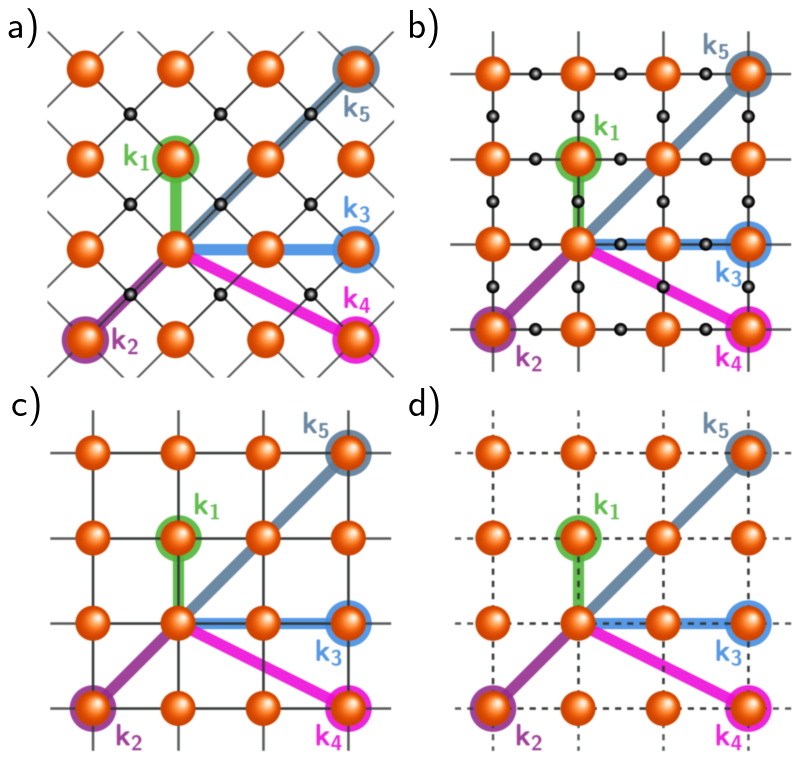}
	\caption{
		The in-plane structures of (a) the 1$n$14, $[M(L)_nM'(L')_4]$, and (b)  1$n$24, $[M(L)_n\{M'(L')_2\}_2]$, families of Hofmann frameworks, (c) the 1$n$02, $[M(L)_n(L')_2]$, family of coordination polymers,  and (d) a simple square supramolecular crystal. In each case the elastic interactions, $k_n$, between $n$th nearest neighbour SCO active $M$ sites (orange circles)  are marked. Non-SCO-active ions ($M'$, black circles) and in-plane ligands ($L'$, black lines) are also shown.   All four classes of materials are described by the same model (Eq. \ref{model}), but with  different magnitudes and signs of the $k_n$ due to the topological differences shown here. 
	}
	\label{lattice} 
\end{figure}

Hofmann-type frameworks that contain one metal species, $ M $, that is SCO active and another, $ M' $, that is diamagnetic have proved a particularly interesting playground for antiferroelasticity (Table \ref{Table:Phases}).\cite{Ni} 
We will discuss two families of Hofmann-type frameworks with the general formulae $[M(L)_nM'(L')_4]$ and $[M(L)_n\{M'(L')_2\}_2]$, where $L'$ is the ligand within the plane, $L$ is the ligand connecting layers, and $n=1$ or $2$ for bridging and monodentate ligands respectively. Henceforth we will refer to these as the 1$n$14 and 1$n$24 families respectively. Prototypical examples are [Fe(pz)Pt(CN)$_4$] and [Fe(pz)\{Au(CN)$_2$\}$_2$]
.\cite{Ni} These crystals have importantly different topologies,  Figs. \ref{lattice}a,b. Nevertheless, in both families the SCO active $M$ sites form simple square sublattices. 

\begin{table*}
	\setlength{\tabcolsep}{3.4pt}
	\begin{tabular}{lccccr} 
		Family & Material & Ref. & $n_{HS}$ plateaus & Phases & Figure \\ 
		\hline
		1$n$14 & \ce{[Fe(thtrz)2Pd(CN)4].EtOH}, \ce{H2O} &  \cite{SciortinoIC,Klein} & 0, $\frac12$, 1 & LS, B, HS & \ref{TwoStepStripe}e \\
		1$n$14 & \ce{\{Fe(Hppy)2[Pd(CN)4]\}.H2O}  &  \cite{Liu17} & 0, $\frac12$, 1 &  LS, B, HS & \ref{TwoStepStripe}e \\
		1$n$14 & \ce{\{Fe(Hppy)2[Pt(CN)4]\}.H2O}  &  \cite{Liu17} & 0, $\frac12$, 1 &  LS, B, HS & \ref{TwoStepStripe}e \\
		1$n$14 & \ce{\{Fe(bpb)[M(CN)4]\}.2naph} &  \cite{lopez} & 0,  $\frac12$, 1 &   LS, B, HS  & \ref{TwoStepStripe}e \\
		1$n$14 & \ce{[Fe(trz-py)2\{Pt(CN)4\}].3H2O} &  \cite{Milin} & $\frac12$, 1 &   B, HS & \ref{TwoStepStripe}d \\
		1$n$14 & \ce{[Fe(proptrz)2Pt(CN)4].2H2O}  &  \cite{Zenere} &  $\frac12$, 1 &   B, HS & \ref{TwoStepStripe}d \\
		1$n$14 & \ce{[Fe(proptrz)2Pd(CN)4].2H2O}  &  \cite{Zenere} &  $\frac12$, 1 &   B, HS & \ref{TwoStepStripe}d \\
		1$n$14 & \ce{[Fe(bztrz)2(Pd(CN)4)].(H2O},\ce{EtOH)} &  \cite{Murphy} & $\frac12$, 1 &   B, HS & \ref{TwoStepStripe}d \\
		1$n$14 & \ce{[Fe(bztrz)2(Pd(CN)4)].3H2O} &  \cite{Murphy} &  $\frac14$, $\frac12$, 1 & X$_\text{L}$,  B, HS & - \\
		1$n$14 & \ce{[Fe(bztrz)2(Pd(CN)4)].$\sim2$H2O} &  \cite{Murphy} & 0, $\frac14$, $\frac12$, 1 & LS, X$_\text{L}$,  B, HS & - \\
		1$n$14 & \ce{[Fe(Hbpt)Pt(CN)4].1/2Hbpt.1/2CH3OH.5/2H2O} &  \cite{Liu18} &  $\frac14$, $\frac12$, $\frac34$, 1 &  X$_\text{L}$, B, X$_\text{H}$, HS & - \\
		1$n$14 & \ce{[Fe(dpsme)Pt(CN)4].2/3 dpsme.$x$EtOH.$y$H2O} &  \cite{Sciortino} & 0, $\frac13$, $\frac12$, 1 & LS, R$_\text{L}$?,  B, HS & \ref{k3Transitions}j \\
		1$n$14 & \ce{[Fe3(saltrz)6(Pt(CN)4)3]8(H2O)} &  \cite{SciortinoCS} & 0, $\frac16$, $\frac23$, $\frac56$, 1 & LS, ??, K$_\text{H}$, ??, HS & S17c \\
		\hline
		1$n$24 & \ce{\{Fe(3-Fpy)2[Au(CN)2]2\}} (form 1; $10^5$ Pa) &  \cite{Agusti_2,Ortega} &  $\frac12$, 1 & A, HS & \ref{TwoStepNeel}d \\
		1$n$24 & \ce{\{Fe(3-Fpy)2[Au(CN)2]2\}} (form 1; 0.18 GPa) &  \cite{Agusti_2,Ortega} & 0, $\frac12$, 1 & LS, A, HS & \ref{TwoStepNeel}f \\
		1$n$24 & \ce{\{Fe(3-Fpy)2[Au(CN)2]2\}} (form 1; 0.26 GPa) &  \cite{Agusti_2,Ortega} & 0, $\frac12$, 1 & LS, A, HS & \ref{TwoStepNeel}g \\
		1$n$24 & \ce{\{Fe(3-Fpy)2[Au(CN)2]2\}} (form 2) &  \cite{Kosone,Ortega} & 0, $\frac12$, 1 & LS, A, HS & \ref{TwoStepNeel}f \\
		1$n$24 & \ce{\{Fe(DMAS)2[Au(CN)2]2\}}  &  \cite{Augusti3} & 0, $\frac12$, 1 & LS, A, HS & \ref{TwoStepNeel}g \\
		1$n$24 & \ce{[Fe(bipytz)(Au(CN)2)2].$x$(EtOH)} &  \cite{Clements} & 0, $\frac13$, $\frac12$, $\frac23$, 1 & LS, D$_\text{L}$, A, D$_\text{H}$, HS & \ref{k3Transitions}d \\
		1$n$24 & \ce{\{Fe(DEAS)2[Ag(CN)2]2\}} &  \cite{Augusti3,Ortega} & $\frac23$, 1 &  D$_\text{H}$, HS & \ref{k3Transitions}c \\
		1$n$24 & \ce{[Fe(isoq)2\{Ag(CN)2\}2]} &  \cite{Meng} & $\frac14$,  $\frac34$, 1 & F$_\text{L}$, F$_\text{H}$, HS & \ref{k5Transitions}a \\
		1$n$24 & \ce{[Fe(dpoda)\{Ag(CN)2\}2].1.5naph} &  \cite{Zhang,Zhang_2} & 0, $\frac14$, $\frac12$, $\frac34$, 1 & LS, F$_\text{L}$, A, F$_\text{H}$?, HS & \ref{k5Transitions}a \\
		1$n$24 & \ce{[Fe(4-abpt)\{Ag(CN)2\}2].2DMF.EtOH} &  \cite{Liu2} & 0, $\frac14$, $\frac12$, $\frac34$, 1 & LS, F$_\text{L}$, A, F$_\text{H}$, HS & \ref{k5Transitions}a \\
		\hline
		1$n$02 & \ce{[Fe(4,4$'$-bipy)2(NCS)2].4CHCl3}	&  \cite{Adams} & 0, $\frac12$, 1 & LS, A, HS & \ref{TwoStepNeel}e \\
		1$n$02 &  \ce{^2{_\infty}[Fe(2,3-bpt)2]}		 &  \cite{bao} & 0, $\frac12$, 1 & LS, A, HS & \ref{TwoStepNeel}g \\
		1$n$02 &  \ce{^2{_\infty}[Fe(2,3-Mebpt)2]}		 &  \cite{bao} & 0, $\frac12$, 1 & LS, A, HS & \ref{TwoStepNeel}g \\
		1$n$02 & \ce{Fe(bpe)2(NCS)2.3(acetone)} &  \cite{Halder2} & 0, $\frac12$, 1 & LS, A, HS & \ref{TwoStepNeel}g \\
		1$n$02 & \ce{Fe2(azpy)4(NCS)4(PrOH)} &  \cite{Halder} & 0, $\frac12$, 1 & LS, A, HS & \ref{TwoStepNeel}g \\
		1$n$02 & \ce{[Fe(bdpt)2]} &  \cite{Lin} & 0, $\frac12$, 1 & LS, A, HS & \ref{TwoStepNeel}e-g$^\dagger$ \\
		1$n$02 & \ce{[Fe(bdpt)2].MeOH} &  \cite{Lin} & 0, $\frac12$, 1 & LS, A, HS & \ref{TwoStepNeel}e-g$^\dagger$ \\
		1$n$02 & \ce{[Fe(bdpt)2].EtOH} &  \cite{Lin} & 0, $\frac12$, 1 & LS, A, HS & \ref{TwoStepNeel}e-g$^\dagger$ \\
		\hline
		molecular & \ce{$\beta$-[FeLBr(dca)2]} &  \cite{Hang} & 0,  $\frac12$, 1 &   LS, B, HS  & \ref{TwoStepStripe}e \\
		molecular & \ce{[Fe(nsal2trien)]SCN} &  \cite{vieira} & 0,  $\frac12$, 1 &   LS, B, HS  & \ref{TwoStepStripe}f \\
		molecular & \ce{[Fe(2-pic)3]Cl2.EtOH} &  \cite{Chernyshov} & 0,  $\frac12$, 1 &   LS, B, HS  & \ref{TwoStepStripe}g \\
		molecular & \ce{[Fe(tdz-py)2(NCS)2]} &  \cite{klingele} & 0,  $\frac12$, 1 &   LS, B, HS  & \ref{TwoStepStripe}f \\
		molecular & \ce{[Mn(3,5-ClSal2(323))]NTf2} &  \cite{Fitzpatrick} & 0,  $\frac12$, 1 &   LS, B, HS  & \ref{TwoStepStripe}f \\
		molecular & \ce{[Fe(salpm)2]ClO4.0.5EtOH} &  \cite{Murnaghan} & 0,  $\frac34$, 1 &   LS,  C$_\text{H}$, HS & \ref{k3Transitions}a,g\\
		molecular & \ce{[Fe(bmpzpy)2][BF4]2.$x$H2O} &  \cite{Money,Ortega} & $\frac12$,  $\frac34$, 1 & B, K$_\text{H}$, HS & \ref{k5Transitions}b\\
		molecular & \ce{[Fe(bpmen)(NCSe)2]} &  \cite{Luan} & 0, $\frac23$, 1 & LS, R$_\text{H}$, HS & \ref{k3Transitions}f \\
		molecular & \ce{[FeH2L^{2Me}]-(PF6)2} &  \cite{Breful,IsingWatanabe} & 0,  $\frac12$, 1 &   LS, S, HS & \ref{k5Transitions}c\\
		molecular & \ce{[Fe(H-5-Brthsa-Me)(5-Br-thsa-Me)].H2O} &  \cite{Li} & 0,  $\frac13$, $\frac23$, 1 &   LS, T$_\text{L}$, T$_\text{H}$, HS & \ref{k5Transitions}d\\
	\end{tabular}
	\caption{Summary of the antiferroelastic order that have been been reported experimentally in the three families of frameworks discussed here and in molecular crystals where the molecules form an approximately square lattice.
		Letters A-T correspond to the phases shown in Fig. \ref{phases}. 
		Subscripts H and L indicate the majority spin state. HS and LS indicate the ferroelastic phases. 
		The figure listed in the rightmost column is one that shows the same qualitative behaviour. Fits to the experimental data have not been attempted.
		A ? after the name of the phase indicates that the assignment is uncertain and a ?? indicates that the ordering is unclear, see the original literature for details.
		The X phase has alternating stripes of width 3 and 1 in  the vertical or horizontal direction (e.g. -LS-LS-LS-HS-LS-LS-LS-HS-; similar to R, which has stripes of width 2 and 1 in  the vertical or horizontal direction, e.g. -LS-LS-HS-LS-LS-HS-). The X phase is not found in the model described here but is found if anisotropy is introduced into the model, for example, by allowing the $k_1$ in the $x$ direction to be different from the $k_1$ in the $y$ direction. 
		$^\dagger$ indicates that the nature of the transitions changes as pressure is applied.
		Here 
		thtrz = \textit{N}-thiophenylidene-4\textit{H}-1,2,4-triazol-4-amine,	
		Hppy = 	4-(1\textit{H}-pyrazol-3-yl)pyridine,
		Hbpt = 4,4’-(1\textit{H}-1,2,4-triazole-3,5-diyl) dipyridine,
		trz-py = 4-(2-pyridyl)-1,2,4,4H-triazole,
		bztrz = (E)-1-phenyl-N-(1,2,4-triazol-4-yl)-methanimine,
		dpsme =	4,4’-di(pyridylthio)methane, 
		MeOH = methanol,
		EtOH = ethanol,
		saltrz = (\textit{E})-2-((((4\textit{H}-1,2,4-triazol-4-yl)imino)methyl)phenol),
		proptrz = (\textit{E})-3-phenyl-\textit{N}-(4\textit{H}-1,2,4-triazol-4-yl)prop-2-yn-1-imine,
		bpb = bis(4-pyridyl)butadiyne,
		naph = naphthalene,
		bipytz = 3,6-bis(4-pyridyl)-1,2,4,5-tetrazine,
		isoq = isoquinoline,
		dpoda = 2,5-di-(pyridyl)-1,3,4-oxadiazole,
		py = pyridine,
		4-abpt=4-amino-3,5-bis(4-pyridyl)-1,2,4-triazole,
		DMAS = 4'-dimethylaminostilbazole,
		DEAS = 4'-diethylaminostilbazole,
		2,3-bptH = 3-(2-pyridyl)-5-(3-pyridyl)-1,2,4-triazole, 
		2,3-MebptH = 3-(3-methyl-2-pyridyl)-5-(3-pyridyl)-1,2,4-triazole,
		bpe = 1,2-bis(4'-pyridyl)ethane),
		azpy = trans-4,4-azopyridine,
		PrOH = propanol,
		bdpt = 3-(5-bromo-2-pyridyl)-5-(4-pyridyl)-1,2,4-triazole,
		tpa = tris(2-pyridylmethyl)amine,
		nsal2trien is obtained by condensation of triethylenetetramine and 2 equiv. of 2-hydroxy-1-naphthaldehyde,
		pic = picolylamine,
		tdz-py = 2,5-di-(2-pyridyl)-1,3,4-thiadiazole,
		salpm = 2-((pyridin-2-ylmethylimino)methyl)phenolate,
		bmpzpy = 2,6-bis{3-methylpyrazol-1-yl}pyridine,
		bpmen = N,N'-dimethyl-N,N'-bis(2-pyridylmethyl)-1,2-ethanediamine,
		\ce{H2L^{2Me}} = bis[N-(2-methylimidazol-4-yl)methylidene-3-aminopropyl]ethylenediamine,
		and H$_2$-5-Br-thsa-Me = 5-bromosalicylaldehyde methylthiosemicarbazone
	}
	\label{Table:Phases}
\end{table*}

Antiferroelasticity has been observed in a number of other coordination polymers where the metal centres form square sublattices, (Fig. \ref{lattice}c; Table \ref{Table:Phases}). For example, the 1$n$02 family of materials with the formula $[M(L)_n(L')_2]$, \textit{e.g.}, [Fe(azpy)$_2$(NCS)$_2$]
(Fig. \ref{lattice}c).\cite{Halder} 

Antiferroelastic order is also found in square lattice supramolecular crystals where molecules are predominantly bound via weak interactions (Fig. \ref{lattice}d; Table \ref{Table:Phases}).\cite{Hang,Wei,Luan,Breful,Matsumoto,Li,Money,Murnaghan,vieira,Chernyshov,klingele,nihei,Fitzpatrick}

Therefore, a thorough theoretical study  of the square lattice combined with a detailed comparison with the extensive experimental literature is an ideal starting point to establish structure-property relations for SCO materials. This is the goal of this paper.

There is a strong correlation between material structure and the collective SCO behaviours (Table \ref{Table:Phases}). A particularly striking example  is the antiferroelastic order in states with  equal numbers of HS and LS ions.
In the 1$n$14 family stripe ordering (Fig. \ref{phases}b) is commonly reported,\cite{Klein,Liu17,Liu18,Milin,Murphy,Sciortino,SciortinoIC,Zenere,lopez} whereas in the 1$n$24\cite{Clements,Zhang,Agusti_2,Zhang_2,Liu2,Augusti3,Kosone}  and 1$n$02\cite{Adams,Halder,Lin,bao,Halder2} families  N\'eel order (also known as checkerboard order, Fig. \ref{phases}a) is  prevalent.

In the 1$n$14 family (Fig. \ref{lattice}a) there are  covalent bonds connecting the second and fifth nearest neighbour  $M$ sites but only weak through-space interactions connecting  nearest, third and fourth neighbours. Whereas, for the 1$n$24 and 1$n$02 families  (Fig. \ref{lattice}b) there are  covalent bonds connecting the nearest and third nearest neighbour $M$ sites and through-space interactions connecting the second, forth and fifth nearest neighbours. 
We will show below that through-bond interactions favour neighbouring metal centres with different spin-states (one high spin and the other low spin); whereas through-space interactions favour neighbouring metal centres with the same spin-states (both high spin or both low spin). This leads to important predicted structure-property relationship for SCO frameworks. For example, this explains why N\'eel order is found in  1$n$24 and 1$n$04 frameworks whereas stripe order is found in 1$n$14 frameworks. A detailed comparison with the experimental literature is given in Table \ref{Table:Phases}. This demonstrates that the antiferroelastic orders previously observed in square lattice SCO materials can be understood from  these structure-property relations or straightforward extensions of them.



\section*{Model}
We start from the free energy difference, $\Delta G_1$, of a single metal centre undergoing spin crossover (SCO):
\begin{equation}
\tilde{\mathcal{H}}_0=\frac{1}{2}\sum_i(\Delta H_1-T\Delta S)\sigma_i\equiv \frac{\Delta G_1}{2}\sum_i \sigma_i,
\end{equation}
where $\Delta H_1=H_H-H_L$ and $\Delta S=S_H-S_L$ are the enthalpy and entropy differences between isolated HS and LS metal centres respectively. Following  Wajnflasz and Pick \cite{WP} we have absorbed the single ion entropy difference into the local Hamiltonian. 
The entropy difference arises from changes in three microscopic terms: the spin and orbital degeneracies are different in the HS and LS states,  and the vibrational entropy also changes due to the softening of the vibrational modes in the HS state. For example, for $\text{Fe}^{2+}$ metal centres in an octahedral complex, $\Delta S_{spin}=k_B \ln 5$ and $\Delta S_{orb}=k_B  \ln 3$.  
Typically, the vibrational contribution is larger, such that $\Delta S\sim 4 \Delta S_{spin}$.\cite{Entropy} Therefore, in all the calculations presented below we set $\Delta S=4k_B\ln5$. $\tilde{\mathcal{H}}_0$ describes spin crossover in non- or weakly-interacting complexes, e.g., in solution. For $\Delta H_1<0$ the HS state is thermodynamically stable at all temperatures. While, for $\Delta H_1>0$ a LS state is realised at low temperatures, gradually undergoing a  crossover to a HS state with equal numbers of HS and LS molecules at $T_{1/2} = \Delta H_1/\Delta S$. 

In the solid state, cooperative elastic interactions between metal centres can lead to first-order phase transitions and hence hysteresis. 
If the neighbouring metal centres are connected through strong covalent bonds then one expects that the potential is close to its minimum and thus the harmonic approximation is reasonable. This justifies modelling the material as a network of springs. 

However, when multiple interactions are present they may be competing such that, it may not be possible to minimize all of the interactions simultaneously -- \textit{i.e.}, there is frustration in the system. Alternatively the interactions may be incommensurate. Paez-Espejo \textit{et al.} \cite{Paez-Espejo}  described this by introducing a `frustration parameter', which measures the extent to which different interactions are minimized by different structures. A limitation with the approach presented by Paez-Espejo \textit{et al.} is that it implicitly assumes the harmonic approximation, which is only valid if all interactions to be near their minima. However, weaker interactions, particularly  through-space  interactions, may be far from their minima.  Here we introduce an approach that removes that difficulty. 

We consider an arbitrary interaction, $V_{ij}(r)$, between two SCO active $M$ sites, $i$ and $j$. 
In a crystal the equilibrium structure minimises the total free energy, which includes the sum of  $V_{ij}(r)$ over all pairs $i,j$. Thus, if the interactions are frustrated there is no guarantee that that any particular $V_{ij}(r)$ is minimised in the equilibrium structure.

The equilibrium separation between nearest neighbour metal centres in the HS phase, $r_H$, is found experimentally to be larger than that in the LS phase, $r_L$.
We
linearly interpolate between these two lengths:
\begin{equation}
r_0 = \overline{R} + \delta(\sigma_i+\sigma_j), \label{eq:r0}
\end{equation}
where $\overline{R}=(r_H+r_L)/2$ and $\delta=(r_H-r_L)/4$. For simplicity we assume that Eq. (\ref{eq:r0}) holds in antiferroelastic phases as well. On noting that $\sigma_i^2=1$ because $\sigma_i=\pm1$, we can write any pairwise symmetric function of the spin-states of the metal ions, $\sigma_i$ and $\sigma_j$, in the form $f(\sigma_i,\sigma_j)=A+B(\sigma_i+\sigma_j)+C\sigma_i\sigma_j$, where $A, B$ and $C$ are constants. Thus, an arbitrary pairwise potential between metal centers can be expanded as
\begin{eqnarray}
\begin{split}
V_{ij}(r,\sigma_i,\sigma_j)=&g_{ij}(r)+h_{ij}(r)\left[r- \eta_{ij}\left\{\overline{R} - \delta(\sigma_i+\sigma_j)\right\}\right]\\
&+\frac{1}{2}k_{ij}(r)\left[r- \eta_{ij}\left\{\overline{R} - \delta(\sigma_i+\sigma_j)\right\}\right]^2,
\end{split}
\end{eqnarray}
where 
$\eta_{ij}=\eta_n=1, \sqrt{2}, 2, \sqrt{5}, 2\sqrt{2}, \dots$ is the ratio of distances between the \textit{n}th and 1st nearest-neighbour distance on the undistorted square lattice.

We interpolate $V_{ij}(r,\sigma_i,\sigma_j)$ by introducing a new function $V_{ij}(r)$ defined such that $V_{ij}(r_H)=V_{ij}(r_H,1,1)$, $V_{ij}(r_L)=V_{ij}(r_L,-1,-1)$, and $V_{ij}(\overline{R})=V_{ij}(\overline{R},1,-1)=V_{ij}(\overline{R},-1,1)$. We show in the supplementary information that 
\begin{eqnarray}
\begin{split}
V_{ij}(r,\sigma_i,\sigma_j)=&f_{ij}(r)+ \delta\eta_{ij} h_{ij}(r)(\sigma_i+\sigma_j)\\
&+\frac{1}{2}k_{ij}(r)\left[r-\overline{R}\eta_{ij} - \delta\eta_{ij}(\sigma_i+\sigma_j)\right]^2;
\end{split}
\end{eqnarray}
and that,   $f_{ij}$, $h_{ij}$, and $k_{ij}$ and independent of $r$ to leading order. Specifically, 
\begin{subequations}
	\begin{eqnarray}
	f_{ij}&=&f_{ij}(\overline{R}\eta_{ij})=V_{ij}(\overline{R}\eta_{ij}),\\
	h_{ij}&=&h_{ij}(\overline{R}\eta_{ij})=V_{ij}^{(1)}(\overline{R}\eta_{ij}), \\
	k_{ij}&=&k_{ij}(\overline{R}\eta_{ij})=V_{ij}^{(2)}(\overline{R}\eta_{ij}),
	\end{eqnarray}
\end{subequations}
where
$V^{(n)}(x)\equiv\left.\left({\partial^n V(r)}/{\partial r^n} \right)\right|_{r=x}.$

The term proportional to $h_{ij}$ has the same functional form as $\tilde{\cal H}_0$ so we define $\Delta H=\Delta H_1 + 4\delta\eta_{ij} \sum_j h_{ij}$ and $\Delta G = \Delta H - T \Delta S$. Thus, we replace $\tilde{\cal H}_0$ by 
\begin{equation}
\mathcal{H}_0=\frac{1}{2}\sum_i(\Delta H-T\Delta S)\sigma_i\equiv \frac{\Delta G}{2}\sum_i \sigma_i.
\end{equation}
This reflects the change in the lattice contribution to the enthalpy when spin states change. However, it does not materially affect the calculations reported here  as we do not calculate $\Delta H$ for specific materials. The term proportional to $k_{ij}$ is simply a harmonic interaction between the metal centres.
Thus, the elastic interactions between metal centres can be represented by a network of springs \textit{even if the individual interactions are far from their minima}.

To investigate spin-state transitions we model the elastic interactions between SCO active sites of the simple square lattice  as a network of springs, illustrated in Fig. \ref{lattice}. We consider elastic interactions between \textit{n}th nearest-neighbour metal centres
\begin{equation}
\mathcal{H}_n=\frac{k_n}{2}\sum_{\langle i,j \rangle_n}\left\{r_{i,j}-\eta_n\left[\overline{R}+\delta(\sigma_i+\sigma_j)\right]\right\}^2
\end{equation}
where $k_n$ is the spring constant between \textit{n}th nearest-neighbour, $\langle i,j \rangle_n$ indicates the sum runs over all \textit{n}th nearest-neighbours, and $r_{i,j}$ is the instantaneous distance between sites $i$ and $j$. 
We include interactions up to \textit{m}th nearest-neighbour metal centres (we study $m=1-5$). Thus, the total Hamiltonian is
\begin{equation}
\mathcal{H}=\sum_{n=0}^m \mathcal{H}_n.
\end{equation}

We solve this model in the `symmetric breathing mode approximation',\cite{Jace} \textit{i.e.}, we assume that for all nearest neighbours, $r_{i,j}=x$, and that the topology of the lattice is not altered by the changes in the spin-states. This yields an effective Ising-Husimi-Temperley model in a longitudinal field
\begin{equation}
\mathcal{H}
\approx\sum_{n=1}^m J_n\sum_{\langle i,j \rangle_n}\sigma_i\sigma_j-\frac{J_\infty}{N}\sum_{i,j}\sigma_i\sigma_j
+\frac{\Delta G}{2}\sum_i\sigma_i,\label{model}
\end{equation}
where, $J_n=k_n \eta_n^2 \delta^2$ is the effective interaction between of the $n$th nearest-neighbour metal centres, 
$J_\infty=\delta^2 \sum_{n=1}^m(k_n z_n \eta_n^2)$ is the long-range strain, $z_n$ is  the coordination number for \textit{n}th nearest neighbours and $N$ is the number of $M$ sites. 
The long-range strain has equal strength between all metal centres, distributing the impact of local molecular volume changes due to spin-state transitions over the lattice. 
Minimization over the instantaneous bond distance requires that $(\partial ^2 \mathcal{H})/(\partial x^2 )=2J_\infty > 0$. Thus the crystal is dynamically unstable for $J_\infty<0$, and  we do not study parameters in that regime below. 


\subsection*{Elastic interactions in materials}

\begin{figure}
	\centering
	\includegraphics[width=0.65\columnwidth]{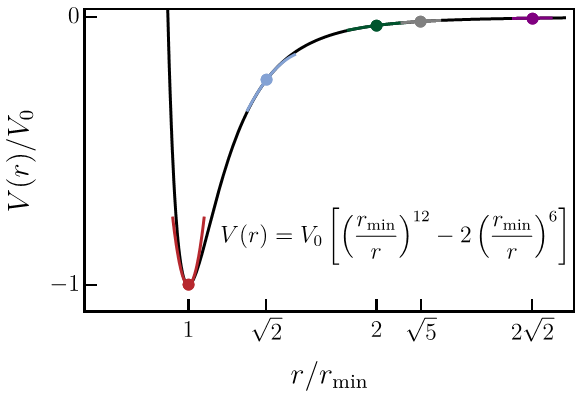}
	\caption{
		The Lennard-Jones potential $V(r)$ between a pair of of molecules separated by a distance $r$. Near the minimum of the potential the second derivative, $V_{ij}^{(2)}(r_0)$, is positive (red curve), thus we expect the elastic interaction, $k_{ij}$, to be strong and positive. Whereas, at larger distances (blue, green, grey and pink curves) the second derivative becomes negative and drops off with increasing distance. Therefore, we expect $k<0$ for through-space interactions away from the potential minimum. 
	}
	\label{LJ} 
\end{figure}

For nearby metal centres joined via (networks of) covalent bonds one expects the metal-metal separation to be close to the minimum of the potential. Hence, one expects that the spring constant, $k_n$, is large and positive (\textit{i.e.}, an antiferroelastic interaction). 
For through-space interactions a separation   larger than the minimum of the potential leads to a negative (ferroelastic) spring constant,  $k_{n}\simeq \partial^2 V_{ij}(r)/\partial r^2|_{r=\overline{R}}$, as illustrated in Fig. \ref{LJ}.

As through-space interactions can be antiferroelastic or ferroelastic, one might find that both $k_1>0$ and $k_2>0$ for some materials in any family of coordination polymers and supramolecular crystals. We will see below that in this case the long-range strain dominates and the SCO transition is always one step.
The three families of frameworks that we consider embody  two distinct topologies with the framework of our model, Fig. \ref{lattice}. In the 1$n$14 family  $k_2$ and $k_5$ are through-bond whereas $k_1$, $k_3$ and $k_4$ are through-space. Therefore,  one expects  $k_2>k_5>0$ and $k_2>|k_1|$, but it is reasonable to expect that in many materials $k_1<k_3<k_4<0$. 
In contrast on the 1$n$24 and 1$n$02 families $k_1$ and $k_3$ are through-bond whereas $k_2$, $k_4$ and $k_5$ are through-space. Therefore,  one expects  $k_1>k_3>0$ and $k_1>|k_2|$, but in many materials one will find that $k_2<k_4<k_5<0$. 
We  show below that these differences are responsible for the different  antiferroelastic orders  observed in the 1$n$14, 1$n$24, and 1$n$02 families.

\section*{Methods}
To construct the zero-temperature phase diagrams we analytically compared the energies of the spin-state phases found in the Monte Carlo calculations (on a variety of lattice sizes) alongside a catalogue of possible phases. 

To investigate the thermodynamic properties of Hamiltonian  we employ Monte Carlo methods using single spin-flips and periodic boundary conditions with $N=60\times 60$ metal ions for the lattices. For each data point we take $N$ measurements with steps of $N$ points after equilibrating for $10N$ steps.

We choose a 60$\times$60 lattice for our Monte Carlo simulations because it is commensurate with every phase observed in the zero-temperature phase diagram for the parameters studied. We will show below that increasing the range of interactions often increases the size of the unit cell for the  antiferroelastic phases and as such, requires a larger grid in our simulations. 

For each parameter set we performed three separate calculations: heating, cooling and parallel tempering. For the heating calculation, we initialize the simulation at the lowest temperature studied ($T=0.01k_1 \delta^2/k_B$) in the $T=0$ ground state predicted by analytic calculations; for higher temperature data points we seed the simulation with the spin-state output from the previous data point. Conversely, for the cooling run we initialize the calculations at the highest temperature studied in a random configuration, then use the resultant output state as a seed for the next data point. When using single spin-flip Monte Carlo the transitions can become frozen out at low temperatures leading to exaggerated predictions of the transition temperatures. We employ parallel tempering to find the lowest free energy state. For the parallel tempering calculations we initialize the simulation in a random configuration.

To calculate the heat capacity, $c_V=c_V^1+c_V^N$, we consider both the single body contribution, $c_V^1$, and the many body contribution, $c_V^N$. The single body contribution comes from the contribution of the the single molecules to the entropy $c_v^1=T(\partial S^1/\partial T)=T\Delta S(\partial n_{HS}/\partial T)$. Where, we have used a Savitzky-Golay filter\cite{SG} to fit $n_{HS}$ and thus calculate $\partial n_{HS}/\partial T$. We calculate the many-body contribution from fluctuations in the true enthalpy $c_V^N=(\langle E^2 \rangle -\langle E \rangle ^2)/(Nk_BT)$, where $E=\mathcal{H}+(1/2)T\Delta S \sum_i\sigma_i$.

\section*{Results and Discussion}

\subsection*{Nearest- and Next Nearest-Neighbour Interactions}


\begin{figure}
	\centering
	\includegraphics[width=0.95\linewidth]{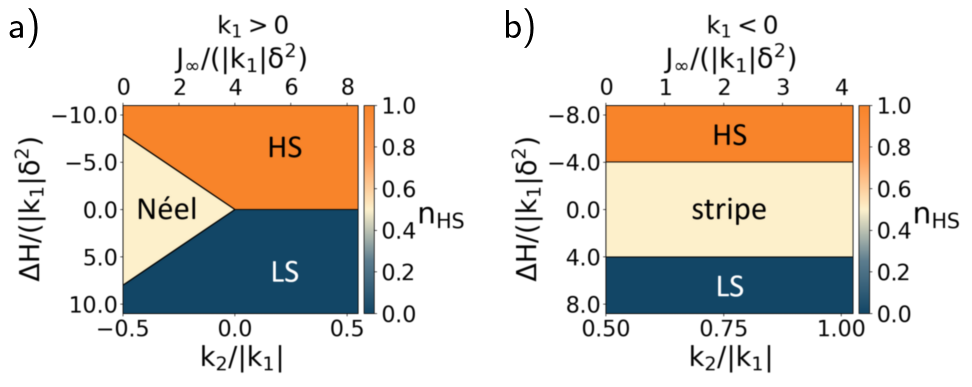} 
	\caption{Zero temperature phase diagrams with nearest ($k_1$) and next nearest ($k_2$) neighbour elastic interactions for (a) the 1$n$24  and 1$n$02 families, and (b) the 1$n$14 family. 
	} 
	\label{k2ZeroTemp} 
\end{figure}

To understand the zero-temperature phase diagrams, Fig. \ref{k2ZeroTemp}, it is helpful to consider which ordering patterns minimise the interactions individually. The  nearest-neighbour interaction is minimised by  N\'eel (also known as checkerboard) order (Fig. \ref{phases}a) for $k_1>0$  and  by HS or LS order, depending on the sign of $\Delta G$, for  $k_1<0$. We will henceforth refer to the HS and LS order as ferroelastic phases. 

The second nearest-neighbour interaction is minimised by stripe order (Fig. \ref{phases}b)  for $k_2>0$ and by either ferroelastic or N\'eel order  for $k_2<0$. 
The elastic interactions cooperate for  $k_1>0$ and $k_2<0$, which are both minimised by N\'eel order (also known as checkerboard order), but are frustrated for any other parameters.

Both the long range strain and the single ion enthalpy favour ferroelasticity. 
These effects compete with the short-range elastic interactions to determine the ground state.  If $J_\infty/(|k_1|\delta^2)$ and $|{\Delta H / (|k_1|\delta^2)}|$ are large then the ground state is either LS or HS,  determined by the sign of $\Delta G$. While if $J_\infty/(|k_1|\delta^2)$ and $|{\Delta H / (|k_1|\delta^2)}|$ are small then either N\'eel  or stripe  ordering is thermodynamically stable. 

\begin{figure*}[h]
	\centering
	\includegraphics[width=0.815\linewidth]{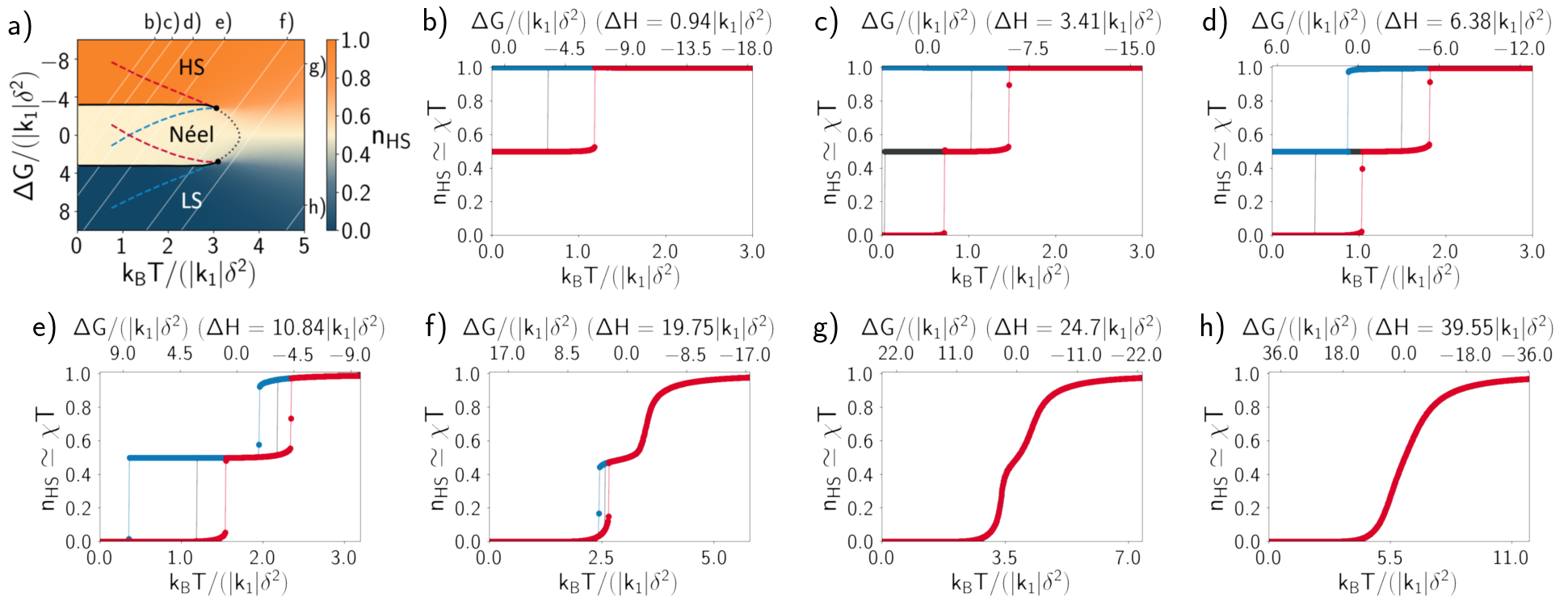} 
	\caption{
		(a) Typical phase diagram for the next nearest neighbour model of the 1$n$24  and 1$n$02 families  with $k_1>0$, $k_2<0$ (here $k_2=-0.2{k_1}$). 
		Shading indicates the fraction of high-spin $M$ sites, $n_{HS}\sim \chi T$ where $\chi$ is the susceptibility, calculated via parallel tempering Monte Carlo. The (black solid) lines of first order transitions become (black dashed) lines of second order transitions  at critical points (black dots).
		The red (blue) dashed lines mark the limits of metastability  cooling (resp. heating), cf. Fig. S5, and hence show  the width of the hysteresis. Individual materials have fixed $\Delta H$,  white lines correspond to panels (b-h), where the fraction of high spins (blue; cooling, red; heating, black; equilibrium) is plotted  (see Fig. S6 for the corresponding  heat capacities).   
	}
	\label{TwoStepNeel} 
\end{figure*}

\subsubsection*{Purely ferroelastic interactions (any family)}

With only nearest neighbour interactions, $k_1>0$, the HS, LS and N\'eel phases are degenerate at $\Delta G=0$ (Fig. \ref{k2ZeroTemp}), and thermal fluctuations stabilise short range N\'eel correlations at elevated temperatures (Fig. S14 and section S2.1). This can lead to a two step transition (Fig. S14d) but does not lead to a spontaneously broken symmetry or long-range order.  Furthermore, the intermediate plateau is observed only in an extremely narrow temperature range. Experimentally,  many different antiferroelastic phases with true long-range order have been found and these phases can be stable over relatively broad temperature ranges. This suggests that longer range elastic interactions are vitally important for multistep transitions.

Considering next nearest neighbour interactions, $k_1>0$ and $k_2>0$ is possible for any of the lattices shown in Fig. \ref{lattice} provided that the minima of both interactions are roughly commensurate with a square lattice.
At $T=0$ 
the LS and HS states are separated by a first order phase transition at $\Delta G=0$, Figs. \ref{k2ZeroTemp}a and S15a. The frustration  entirely suppresses the N\'eel order found at finite temperatures for $k_2=0$ (cf. Figs. S14 and S15).

For constant $\Delta H$, which represents individual materials, we find a single step transition (Fig. S15). Which can be sharp and first order, second order, or a crossover, depending on the relative strengths of the elastic interactions and the single ion entropy (see section S23.2).
Consistent with this prediction, single step transitions are common and observed in all of the families of materials discussed here.

\subsubsection*{Antiferroelastic interactions in the 1$n$24  and 1$n$02 families}

One  expects that many materials in 1$n$24  and 1$n$02 families will have ferroelastic next nearest neighbour interactions  ($k_2<0$, which requires $k_1>2|k_2|$). This leads to a much richer range of behaviours, Fig. \ref{TwoStepNeel}. The N\'eel phase is stable at $T=0$ and there are two lines of first order transitions ending at two critical points joined by a line of second order transitions, Fig. \ref{TwoStepNeel}a. For individual materials with fixed $\Delta H$ this leads to seven thermodynamically distinct behaviours. 
Generically, there is a two step transition from HS to N\'eel to LS as the temperature is lowered. Each step can be either a crossover, a first order transition or a second order transition. 

If the  single ion enthalpies of the HS and LS states are finely balanced (small $|\Delta H|/(|k|\delta^2)$) then there is a first order, one step, incomplete transition  between the N\'eel ordered phase and the HS phase (Figs. \ref{TwoStepNeel}b-c and S6b-c). An incomplete one-step transition is even observed when $\Delta H$ is small and negative, Figs. \ref{TwoStepNeel}b and S6b. This is remarkable as the single ion free energy, $\Delta G$, favours the HS state at all temperatures. This transition is a truely collective effect driven by the system's need to minimise the energy of the elastic interactions, which are strong in this regime, in order to minimise the total free energy of the system. 

In some cases the hysteresis is sufficiently broad that straightforwardly cooling the system does not achieve the true low temperature ground state. This is observed experimentally and has been called `hidden hysteresis'.\cite{Milin2} Nevertheless it may be possible to prepare the ground state either via the reverse LIESST effect or by applying and subsequently adiabatically releasing pressure. 

For larger  $\Delta G/(|k|\delta^2)$ both steps are straightforwardly observable and show significant hysteresis (Figs. \ref{TwoStepNeel}e and S6e). The width of the hysteresis loops decrease as $\Delta G/(|k|\delta^2)$ increases. The low temperature step always displays wider hysteresis than the high temperature step. On further increasing $\Delta G/(|k|\delta^2)$ first the high temperature step passes through the critical point and becomes second order, with the low temperature step remaining first order and hysteretic (Figs. \ref{TwoStepNeel}f and S6f). As $\Delta H$ is further increased the lower temperature step passes through the critical point and both transitions become crossovers (Figs. \ref{TwoStepNeel}g and S6g). There are no parameters for which the high temperature step is first order and the low temperature step is a second order (cf. Fig. \ref{TwoStepNeel}a). For sufficiently large $\Delta G/(|k|\delta^2)$ the elastic interactions become unimportant and the temperature dependence of $n_{HS}$ begins to resemble a single crossover (Figs. \ref{TwoStepNeel}h and S6h).  The distinction between a second order transition and a crossover regimes is much clearer in the heat capacity, Fig. S6, than in $n_{HS}$. 
\footnote{This is because $n_{HS}$ is a first derivative of the free energy, whereas heat capacity is a second derivative of the free energy. Therefore, $\partial n_{HS}/\partial T$ contains similar information to the heat capacity.}


In all of these cases, the phase with $n_{HS}\simeq1/2$ is N\'eel ordered. The N\'eel phase is found experimentally in many materials in the 1$n$24\cite{Clements,Zhang,Agusti_2,Zhang_2,Liu2,Augusti3,Kosone} and 1$n$02\cite{bao,Halder2,Adams,Halder,Lin} families, and also in supramolecular crystals\cite{Wei,nihei} (see Table \ref{Table:Phases}).  Just as we report here,  N\'eel ordering is observed experimentally both as the intermediate spin-state in two-step  transitions,\cite{Agusti_2,Augusti3,Kosone,bao,Halder2,Adams,Lin,Wei}  and as the low temperature phase in incomplete one-step transitions \cite{Agusti_2,Augusti3,Adams,Halder}.

\subsubsection*{Antiferroelastic interactions in the 1$n$14  family}

\begin{figure*}
	\centering
	\includegraphics[width=0.815\linewidth]{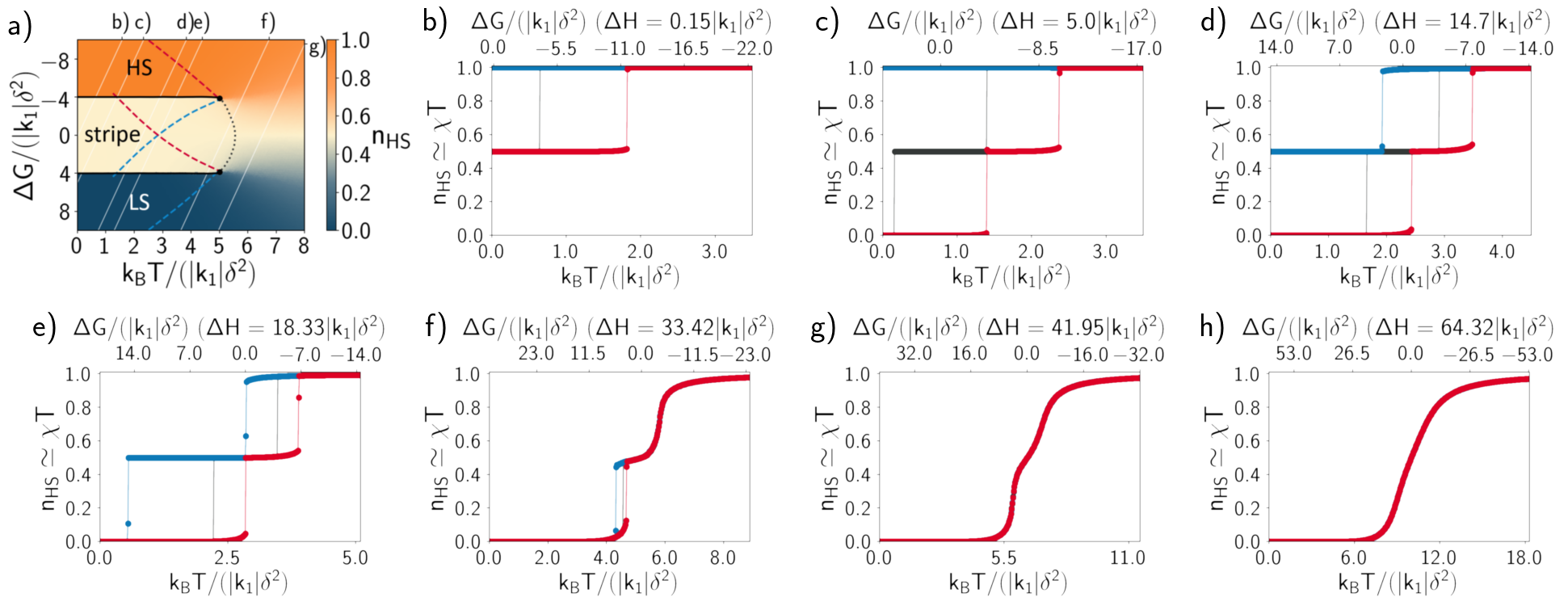} 
	\caption{
		(a) Typical phase diagram for the next nearest neighbour model of the 1$n$14 family with $k_1<0$ and $k_2>0$ (here $k_2=1.2|k_1|$; see also Fig. S7).  (b-f) The fraction of high spins, $n_{HS}$ (see Fig. S8 for the corresponding heat capacities). Symbols have the same meanings as in Fig. \ref{TwoStepNeel}. 
		In contrast to the  1$n$24 and 1$n$02 families (Fig. \ref{TwoStepNeel}) the antiferroelastic order is striped (Fig. \ref{phases}b) rather than N\'eel (Fig. \ref{phases}a). Otherwise the behaviours are extremely similar.
	}
	\label{TwoStepStripe} 
\end{figure*}

One expects that for many 1$n$14  materials $k_1<0$ and $k_2>|k_1|/2$. This leads to antiferroelastic states with stripe order (Figs. \ref{k2ZeroTemp}b, \ref{TwoStepStripe} and S13).  Apart from this important difference, the thermodynamic behaviour is extremely similar to that predicted for the 1$n$24  and 1$n$02 families ($k_1>0$, $k_2<0$; Figs. \ref{TwoStepNeel} and S6). Both the N\'eel and stripe phases have $n_{HS}=1/2$. In both cases we observe two lines of first order transitions ending in two critical points (compare Fig. \ref{TwoStepStripe}a to \ref{TwoStepNeel}a). This gives rise to the same range of possible behaviours for individual materials (fixed $\Delta H$) as we found on the 1$n$24 family and 1$n$02 families: compare Figs. \ref{TwoStepNeel} and  S6 to  \ref{TwoStepStripe} and S8. In contrast to the purely ferroelastic case ($k_1>0$ and $k_2>0$) when one of the interactions is antiferroelastic the relative magnitudes $k_1$ and $k_2$ has an important consequences for the phases diagram, must notably how stable the antiferroelastic phase is, compare Figs. \ref{TwoStepStripe} and S13.

Stripe order has been observed in many 1$n$14  frameworks\cite{Klein,Liu17,Liu18,Milin,Murphy,Sciortino,SciortinoIC,Zenere,lopez} and also in supramolecular crystals\cite{Hang,Matsumoto,vieira,Chernyshov,klingele,Fitzpatrick}, see Table \ref{Table:Phases}. Just as we find in our calculations stripe ordering is found as the low temperature phase in an incomplete first-order transition\cite{Zenere,Milin,Murphy} and as the intermediate phase in two-step transitions\cite{Milin,SciortinoIC,Liu17,Klein,lopez,Hang,vieira,Chernyshov,klingele,Fitzpatrick}.

N\'eel ordering is  commonly found as an intermediate spin-state in the 1$n$24  and 1$n$02 families  while, stripe ordering is  common in the 1$n$14 family. Our calculations therefore give a clear explanation for this structure-property relationship. In the 1$n$24  and 1$n$02 families  one expects the $k_1$ interactions to be strong and antiferroelastic (positive) as they are through-bond, but the $k_2$ interactions to be weaker and possibly ferroelastic (negative) as they are through-space -- this favours N\'eel order. Conversely, in a 1$n$14 family  one expects the $k_1$ interactions to be weaker and possibly ferroelastic (negative) as they are through-space, but the $k_2$ interactions to be strong and antiferroelastic (positive) as they are through-bond -- this favours stripe order.

Phase diagrams for two-step transition have previously been reported for a phenomenological Ising-like model\cite{IsingWatanabe} and a Landau theory.\cite{IsingChern} Many of the same qualitative features can be observed. 

\subsection*{Third Nearest-Neighbour Interactions}

The third nearest-neighbour interaction, $k_3$, is through-bond in the 1$n$24  and 1$n$02 families, but through-space in the 1$n$14 family, Fig. \ref{lattice}. We therefore expect that it will be antiferroelastic ($k_3>0$) for most materials in the 1$n$24  and 1$n$02 families  but it may be ferroelastic ($k_3<0$) for members of the 1$n$14 family. Thus $k_3$ could have significantly different effects on the two different classes of materials.
In supramolecular crystals for non-zero $k_1$, $k_2$ and $k_3$ any one or any pair of elastic constants are may be negative  so long as the lattice remains dynamically stable ($J_\infty=4k_1+8k_2+16k_3>0$). 


Interestingly, in addition to the phases that minimize the energy of any single elastic interaction, elastic frustration  introduces additional phases into the zero-temperature phase diagram, Fig. S16. We find 13 states that are thermodynamically stable at $T=0$ in extended regions of  parameter space: HS, LS, N\'eel, stripe, C$_H$, C$_L$, D$_H$, D$_L$, E, G, R$_H$, R$_L$ and S, cf. Fig. \ref{phases}, where the subscript indicates the majority spin-state.  

\begin{figure}
	\centering
	\includegraphics[width=0.95\linewidth]{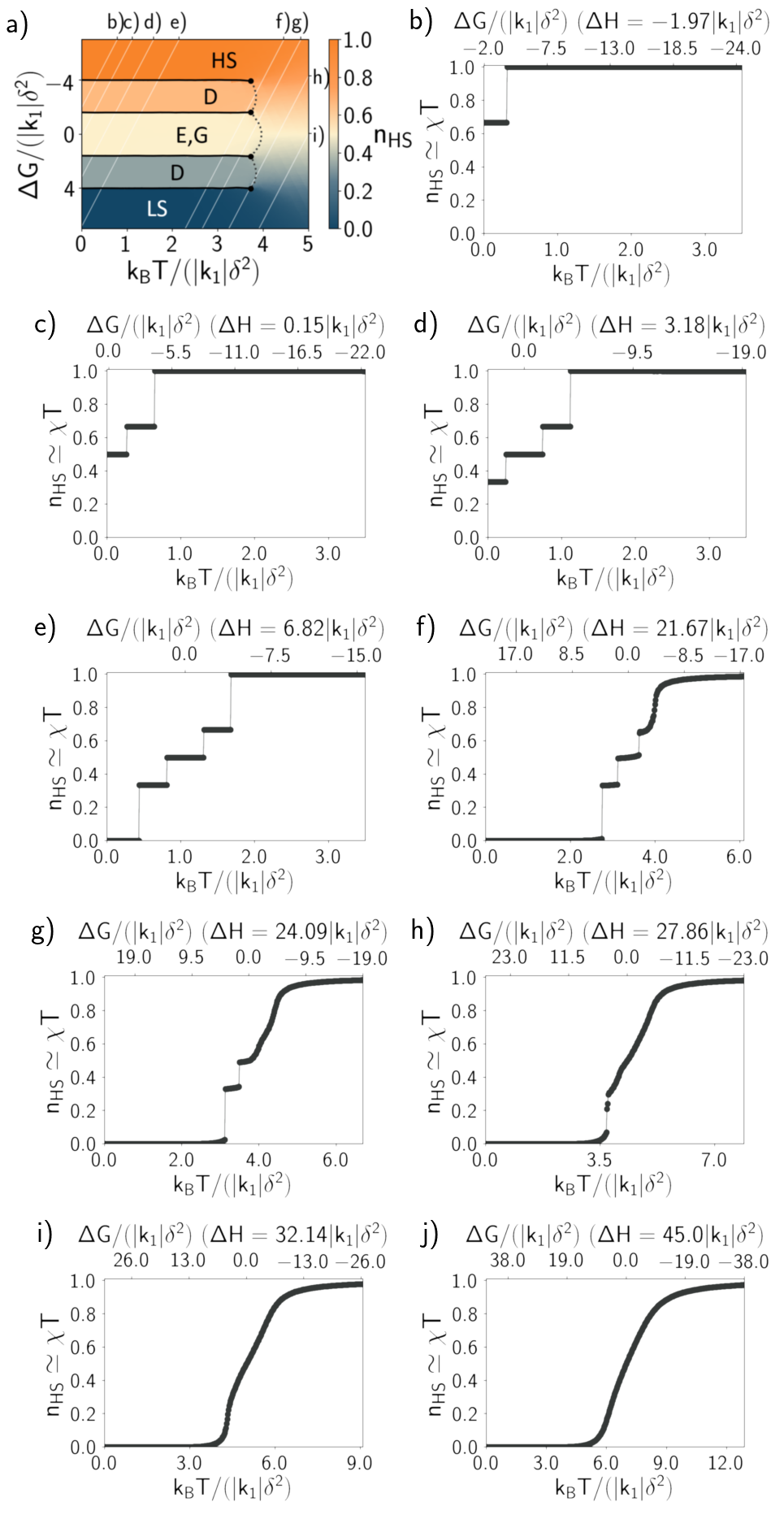} 
	\caption{
		(a) Typical slice of the phase diagram for the third nearest neighbour model of the 1$n$24  and 1$n$02 families with $k_1$ and $k_3>0$, and $k_2<0$ (here  $k_2=-0.9k_1$ and $k_3=0.5k_1$). (b-i) The fraction of high spins, $n_{HS}$ (see Fig. S9 for the corresponding heat capacities). For simplicity we only show the   parallel tempering Monte Carlo predictions.  Symbols have the same meanings as in Fig. \ref{TwoStepNeel}.
	}
	\label{FourStepB} 
\end{figure}

\begin{figure}
	\centering
	\includegraphics[width=0.95\linewidth]{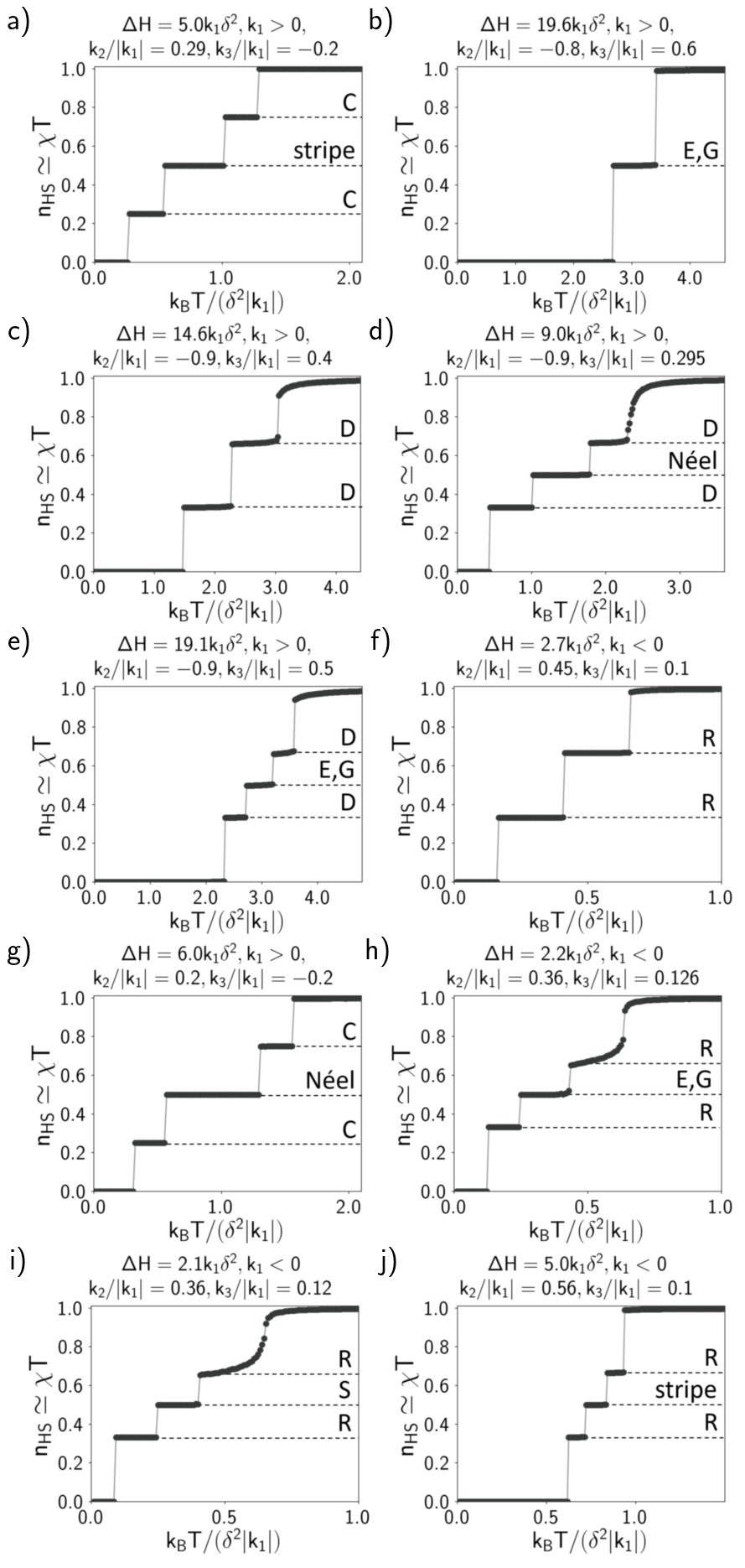} 
	\caption{Examples of the wide range of behaviours found in the third nearest neighbour model. Here we study  parameters relevant to  (a) the 1$n$14 family, (b-e) the 1$n$24  and 1$n$02 families and (a-j) supramolecular lattices. Only the parallel tempering Monte Carlo simulations are shown. The intermediate spin-state phases are labelled on the plots (cf. Fig. \ref{phases}). The corresponding heat capacities are shown in Fig. S10.
	}
	\label{k3Transitions} 
\end{figure}

For materials in the 1$n$24  and 1$n$02 families  one expects $k_1>0$ and $k_3>0$, and in many materials one may find $k_2<0$. A typical slice of the finite temperature phase diagram for this parameter regime is shown in Fig. \ref{FourStepB}a. This phase diagram has stable states consistent with plateaus at $n_{HS}= 1$ (HS), $\frac23$ (D$_H$), $\frac12$ (E or G, which are degenerate), $\frac13$ (D$_L$), or $0$ (LS). Here we have chosen an example where the third nearest neighbour interaction has changed the order at $n_{HS}=\frac12$. But, note that, for other parameters we find similar phase diagrams where the N\'eel order remains (cf. Figs. \ref{k3Transitions}d,g and S16)

Considering  lines of constant $\Delta H$ corresponding to individual materials,  Figs. \ref{FourStepB}b-j and S9b-j, we observe a rich range of behaviours. For small $\Delta H$ we find incomplete one-, two-, and three-step  transitions (Figs. \ref{FourStepB}b-d and S9b-d), where all the transitions are first order. Note that again we find incomplete transitions driven purely by the elastic interactions with $\Delta H<0$, \textit{i.e.}, when the the single ion free energy favours  HS ions at all temperatures.
For moderate $\Delta H$ a complete four step transition is observed as four first order transitions  (Figs. \ref{FourStepB}e and S9e). As $\Delta H$ is increased the high temperature transitions  successively  become second order and then crossovers  (Figs. \ref{FourStepB}e-j and S9e-j). Interestingly, for large $\Delta H$ the E or G phase ($n_{HS}= \frac12$) is suppressed by thermal fluctuations and we see either one crossover and two phase transitions (one first and one second order; Figs. \ref{FourStepB}h and S9h) or a single crossover where the remains a sharp drop in $n_{HS}$ at low temperature, but no hysteresis (Figs. \ref{FourStepB}i and S9i). Eventually, for large $\Delta H$ the effects of the elastic interactions are negligible and there is a single smooth crossover from HS to LS  (Figs. \ref{FourStepB}j and S9j). The increased complexity caused by these multistep transitions means that the heat capacity (Fig. S9) and $\partial n_{HS}/\partial T$ are more sensitive probes of the number and location of the transitions/crossovers. Alternatively, one could employ the techniques that are directly sensitive to spatial symmetry breaking that accompanies the development of long-range spin-state order; for example, x-ray scattering (see Fig. \ref{phases}). However, this  requires monitoring the changes in the Bragg peaks as temperature varies\cite{IsingChern,IsingWatanabe} -- and such experiments are not commonplace.

While Fig. \ref{FourStepB} is typical of the rich behaviour of the third nearest neighbour model, the behaviours displayed there are far from exhaustive of this  model. In Figs. \ref{k3Transitions}b-e (and Figs. S10b-e) we show some additional examples of two-, three-, and four-step transitions  with $k_1>0$, $k_2<0$ and $k_3>0$ (relevant to the  1$n$24  and 1$n$02 families). Importantly the antiferroelastic order can vary with the N\'eel phase competing with the E and G phases at $n_{HS}=1/2$.  D phases are often stable for $n_{HS}=1/3$ or $2/3$ in this parameter regime. Although we only show a single temperature trace for each case we emphasise that for each case varying $\Delta H$ results in the same range of thermodynamic behaviours as we saw in Fig. \ref{FourStepB}, with incomplete transitions, first order transitions, second order transitions, and crossovers observed. 

Some of the phenomenology found here has been reported in experimental literature, for example Clements \textit{et al.}\cite{Clements} reported a four-step transition with intermediate plateaus at $n_{HS}= 2/3$, $1/2$, and $1/3$ in [Fe(bipytz)(Au(CN)$_2$)$_2$]$\cdot x$(EtOH), which is in the 1$n$24 family. 
They demonstrated that the antiferroelastic order in these plateaus is D$_H$, N\'eel, and D$_L$ respectively. This is precisely the behaviour shown in Fig. \ref{k3Transitions}d (and Fig. S10d). Table \ref{Table:Phases} shows that similar agreement is found with many other experiments.

For the 1$n$14 family  one expects $k_2>0$ and, in many materials, $k_1<0$ and $k_3<0$. In this regime we observe only one- and two-step transitions. The latter with intermediate stripe order similar to the transitions reported in Fig. \ref{TwoStepStripe}. However,  one also expects that in some materials $k_1>0$. In this regime we find a four-step transition with intermediate $n_{HS}= 1/4$ (C$_L$), $1/2$ (stripe) and $3/4$ (C$_H$) plateaus (Figs. \ref{k3Transitions}a and S10a).

Many materials in the all families will have $k_1>0$, $k_2>0$, and  $k_3>0$. In this case the long-range strain ($J_\infty$) dominates over the elastic interactions. This allows only one-step transitions. 

For supramolecular crystals the only constraint on the parameters is that the lattice is stable ($i.e.$, that $J_\infty=4(k_1+2k_2+4k_3)>0$). This allows for an even greater range of possibilities. Some of these are demonstrated in Figs. \ref{k3Transitions}a-j and S10a-j. We see four-step transitions with a variety of different antiferroelastic orders and fractions of HS in the intermediate plateaus.
Similar to the previous cases studied, varying $\Delta H/ (k_1\delta^2)$ can lead to incomplete transitions, first order transitions, second order transitions and crossovers.  Several of these behaviours have been reported in the experimental literature. For example,  two step transitions with  intermediate stripe phases have been reported by Hang \textit{et al.}\cite{Hang}, Vieira \textit{et al.}\cite{vieira}, Chernyshov \textit{et al.}\cite{Chernyshov}, Klingele \textit{et al.}\cite{klingele}, Fitzpatrick \textit{et al.}\cite{Fitzpatrick} and a two-step incomplete transition with a low temperature stripe and intermediate C$_H$  ($n_{HS}=0.75$) antiferroelastic states has been reported by Matasumoto \textit{et al.}\cite{Matsumoto}. 


Overall the third nearest neighbour model suggests that longer range elastic interactions lead to a wider range of  behaviours. This includes transitions with a greater number of steps and a greater variety of antiferroelastic order.

\subsection*{Longer Range Interactions}

Forth nearest neighbour interactions are through-space in  the 1$n$14, 1$n$24  and 1$n$02 families, thus one might expect it to be weak in all classes of materials.
However, $k_5$ is through-space in the 1$n$24  and 1$n$02 families but through-bond in the 1$n$14 family. Thus, we expect $k_5$ to be positive in  1$n$14 family materials. Therefore, to test our proposal that longer range interactions generically increase the number of plateaus in SCO materials we study our model with $k_4=0$ and non-zero  $k_1$, $k_2$, $k_3$ and $k_5$. 

For materials in the 1$n$14 family  one expects $k_2>0$ and $k_5>0$, but for many materials  $k_1<0$ and $k_3<0$. 
This leads to stable states consistent with plateaus at $n_{HS}= 1$ (HS), $\frac23$ (R$_H$), $\frac12$ (stripe), $\frac13$ (R$_L$), or $0$ (LS), Fig. S18a. Other than the change in the antiferroelastic order the behaviours are extremely similar to those shown in Fig. \ref{FourStepB} where we also find plateaus at the same HS fractions. The five phases are separated by  four first order lines ending at critical points where a line of second order transitions begin. Once again, for smaller values of $\Delta H/(|k_1|\delta^2)$ the transitions are sharp and first order, broadening for increasing values until the transitions become second order and then crossovers (see Figs. S18b-j and S11b-j).

\begin{figure}
	\centering
	\includegraphics[width=0.95\linewidth]{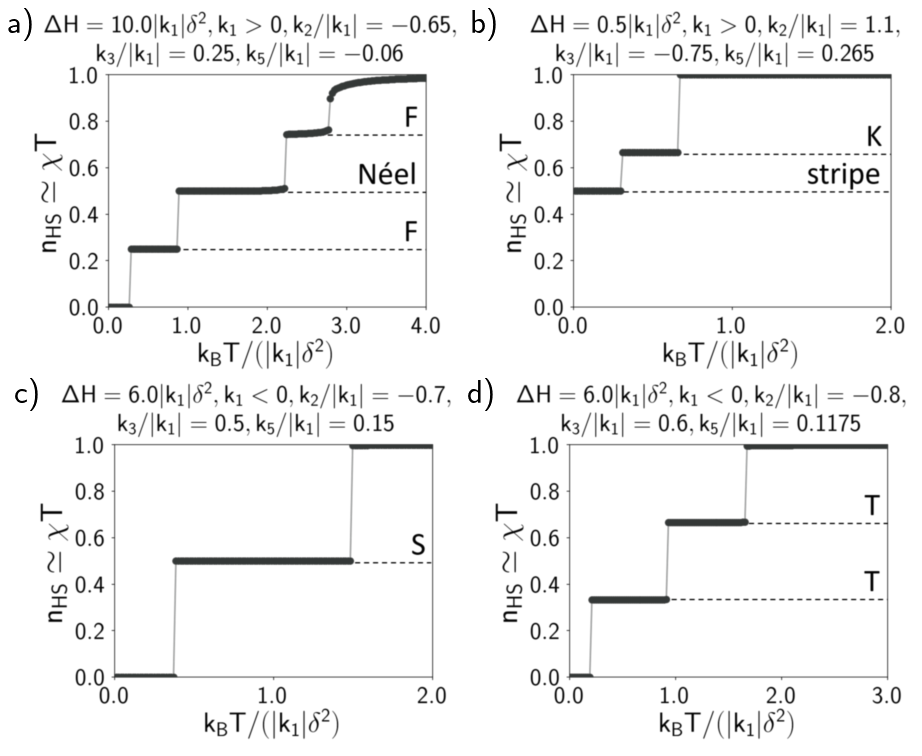} 
	\caption{Examples of the wide range of behaviours found in the fifth nearest neighbour model. Here we have selected behaviours observed experimentally, see  Table \ref{Table:Phases}. The intermediate spin-state phases are labelled on the plots (cf. Fig. \ref{phases}). 
	}
	\label{k5Transitions} 
\end{figure}

The inclusion of the $k_5$ interaction also allows for a large number of possible antiferroelastic states and an extremely rich phase diagram, Fig. S17. At $T=0$ we have identified 36 possible ground states that are stabilised in extended regions of the phase diagram, Fig. \ref{phases}. For parameters relevant to 1$n$14 family ($k_2>0$ and $k_5>0$) the  HS, LS, stripe (B), C, J, K, R, S, and I phases are predicted. For parameters relevant to 1$n$24  and 1$n$02 families ($k_1>0$ and $k_3>0$)  the  HS, LS, N\'eel (A), stripe (B), C, D, E, F, G, L, M, N, O, and P phases are found. Once again, this is consistent with experimental literature (Table \ref{Table:Phases}), where K\cite{SciortinoCS} and R\cite{Sciortino} phases have been reported for materials in the 1$n$14 family;  and D\cite{Clements,Augusti3} and F\cite{Meng,Zhang,Zhang_2,Liu2} phases have been observed in materials in the 1$n$24 family.
A few examples of the possible  dependence of $n_{HS}$ as one sweeps the temperature are given in Fig. \ref{k5Transitions}. These examples have been selected for their relevance to the existing experimental literature, cf. Table \ref{Table:Phases}, and are far from exhaustive.

If anisotropy is added to the model -- either via crystallographically distinct metal centers or through anisotropic interactions, e.g., allowing $k_1$ to be different in the $x$ and $y$ directions -- then even more phases are found. A particularly important example in the X phase, which consists of alternating stripes of width 3 and 1 in  the vertical or horizontal direction, e.g. -LS-LS-LS-HS-LS-LS-LS-HS-. This is similar to the R phase (Fig. \ref{phases}), which has stripes of width 2 and 1 in  the vertical or horizontal direction (-LS-LS-HS-LS-LS-HS-). The fact that the X phase is observed in several members of the 1$n$24 family (Table \ref{Table:Phases}) suggests that the interplay between anisotropy and elastic interactions may play an important role in determining which spin-state orders are observed.

Thus, most of the phenomenology reported in experimental literature is thus reproduced in our calculations  (see Table \ref{Table:Phases} and Fig. \ref{k5Transitions}). For example four step transitions with F$_\text{L}$ ($n_{HS}=0.25$), A ($0.5$), and F$_\text{H}$ ($0.75$) phases (Fig. \ref{k5Transitions}a) have been reported in  1$n$24 frameworks,\cite{Zhang,Zhang_2,Liu2}; whereas in molecular crystals there have been observations of two step transitions with intermediate stripe (B; $n_{HS}=0.5$)  and $\text{K}_{\text{H}}$ (0.75) phases \cite{Money,Ortega} (Fig. \ref{k5Transitions}b), two step transitions with an intermediate S (0.5) phase\cite{Breful,IsingWatanabe} (Fig. \ref{k5Transitions}c), and a three step transition with intermediate   T$_\text{L}$ ($\frac13$) and T$_\text{H}$ ($\frac23$) phases\cite{Li} (Fig. \ref{k5Transitions}d).

\begin{figure}
	\centering
	\includegraphics[width=0.95\linewidth]{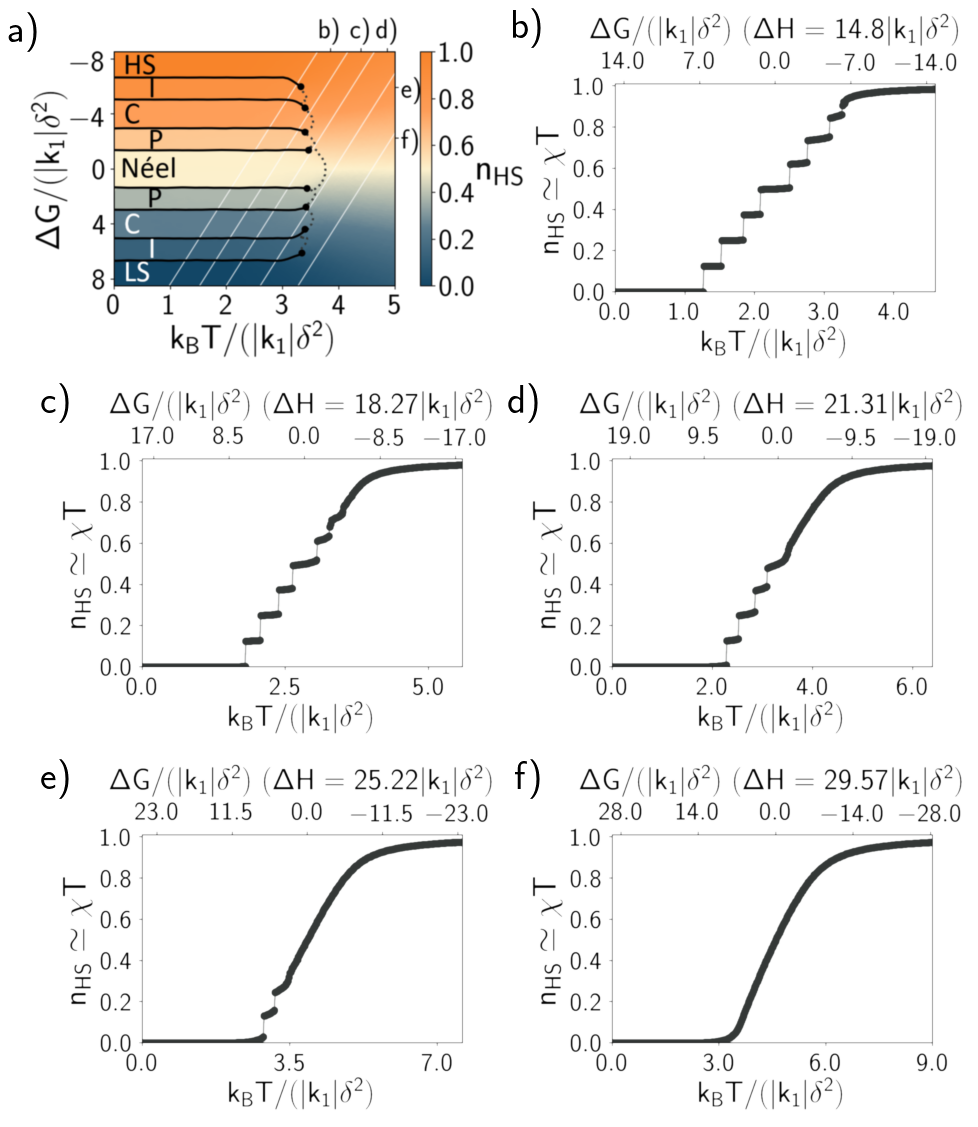} 
	\caption{(a) Typical slice of the finite temperature phase diagram for  interactions up to fifth nearest neighbours as expected in the 1$n$24 and 1$n$02 families with $k_1$, $k_2$ and $k_3>0$, and $k_5<0$ (here $k_2=0.3k_1$, $k_3=0.07k_1$ and $k_5=-0.16k_1$). Lines and dots have the same meanings as in Fig. \ref{TwoStepNeel}a.
		(b-f) The fraction of high spins, $n_{HS}$ (See Fig. S12 for the corresponding heat capacities). Only the parallel tempering predictions are shown.}
	\label{EightSteps} 
\end{figure}

The diversity of the phases, Fig. \ref{EightSteps}a, also leads to a very large numbers of steps. For example,  in Fig. \ref{EightSteps}b (and Fig. S12b) we report an eight-step transition, for parameters relevant to 1$n$24  and 1$n$02 families. In the final stages of preparing this manuscript we became aware of an eight step transition reported recently by Peng \textit{et al.}\cite{Peng} for a 1$n$24  material. 

As in the  cases described above, increasing $\Delta H/(|k_1|\delta^2)$ increases the widths of the transitions, and eventually causes them to become second order and then crossovers. This starts with the highest temperature transitions and works progressively down (see Figs. \ref{EightSteps}b-e). For large $\Delta H/(|k_1|\delta^2)$ the crossover becomes extremely smooth as it passes through all nine phases (see Figs. \ref{EightSteps}f and S12f). Thus, $n_{HS}$, does not show any significant differences from a trivial single crossover. However, clear signatures are still observed in the $c_V$ and $\partial n_{HS}/\partial T$, see Fig. S12. Unless the heat capacity, $\partial n_{HS}/\partial T$, or  temperature dependent x-ray scattering are measured then this highly-multi-step crossover could be broadly dismissed as a single broad crossover. This means that extremely multistep SCO crossovers may hidden in plain sight in the literature. It is interesting to note that while, in this two dimensional model increasing the range of interactions on these two-dimensional lattices leads to transitions with an increasing numbers of steps, in the one dimensional Ising model only next nearest interactions are required to see an infinite number of steps (Devil's staircase)\cite{Devil,DevilExp}. In contrast one does not expect the model in three dimensions to be dramatically different from the two dimensional model studied here, although that may be an interesting avenue for future study as it would allow a full understanding of the interlayer ordering in SCO materials.

\section*{Conclusions}

The simple elastic model explored here hosts a rich variety of SCO transitions and intermediate spin-state phases. Almost all of the spin-state ordered phases  reported experimentally in square lattice SCO materials are found in our calculations, see Table \ref{Table:Phases}. However, we also predict new orderings that have not yet been observed, see Fig. \ref{phases}.

We have established clear structure-property relations for the  1$n$14, 1$n$24 and 1$n$02 families of SCO materials. The key point is that through-bond elastic interactions are generically antiferroelastic (described by positive spring constants) whereas through-space elastic interactions can often be ferroelastic (negative spring constants). This provides a natural explanation for the different antiferroelastic phases found in different families of SCO frameworks. 

In general, increasing the range of  interactions results in an increase in the number of observed transitions and spin-state phases. Therefore, the  rigidity of framework materials  may explain why they are such a rich playground for multistep transitions and antiferroelastic order. 

Our results show that multistep transitions and complex patterns of antiferroelastic order on lattices with initially equivalent metal ions can be understood in terms of the competition and cooperation between through-bond and through-space  interactions. The range of the elastic interactions is a vital criteria for understanding collective phenomena in SCO materials. 
Our results show  that strong through-space interactions are the key requirement for multi-step transitions. This explains why and how the presence of guest molecules, solvent atoms and anions \cite{ColletReview} can strongly modify collective SCO behaviours -- through hydrogen bonding, aromatic $\pi$-$\pi$ bonding and van der Waals interactions. 

The inclusion of up to fifth nearest neighbour elastic interactions allows for eight-step transitions for parameters relevant to the 1$n$24  and 1$n$02 families. This suggests that transitions with larger numbers of steps  could be obtained more readily in the 1$n$24  and 1$n$02 families  than in the 1$n$14 family.

Due to the abundance of experimental literature on the 1$n$24, 1$n$02 and 1$n$04 families of frameworks we have specialised, in this paper, to studying the square lattice. However, the same theory presented here is applicable to understand SCO materials on any lattice.\cite{Jace,JacePyro} 

Further insight into these trends could be gained by parametrising our model for specific materials. However, this remains a major challenge.\cite{Miriam}

\section*{Conflicts of interest}
There are no conflicts to declare.

\section*{Acknowledgements}
We thank Cameron Kepert, Ross McKenzie, Suzanne Neville, and Gian Ruzzi for their helpful conversations. This work was funded by an Australian Government Research Training Program Scholarship.  This work was supported by the Australian Research Council through grant number DP200100305.



\balance



\clearpage

\onecolumn

\renewcommand\thefigure{S\arabic{figure}}
\renewcommand\thesection{S\arabic{section}}
\renewcommand\thesubsection{S\arabic{section}.\arabic{subsection}}
\setcounter{figure}{0} 
\setcounter{equation}{0} 


\begin{LARGE}
	\begin{center}
		\textbf{SUPPLEMENTARY INFORMATION}
	\end{center}
\end{LARGE}
	
\section{Expansion of the potential}

In the main text  (Eq. 3) we consider a pairwise potential between neighbouring metal cites that depends on the spin spin-states of the metal ions, $\sigma_i$ and $\sigma_j$. This potential is given by
\begin{eqnarray}
\begin{split}
V_{ij}(r,\sigma_i,\sigma_j)=&g_{ij}(r)+h_{ij}(r)\left[r- \eta_{ij}\left\{\overline{R} - \delta(\sigma_i+\sigma_j)\right\}\right]
+\frac{1}{2}k_{ij}(r)\left[r- \eta_{ij}\left\{\overline{R} - \delta(\sigma_i+\sigma_j)\right\}\right]^2,
\end{split}
\end{eqnarray}
where 
$\eta_{ij}=\eta_n=1, \sqrt{2}, 2, \sqrt{5}, 2\sqrt{2}, \dots$ is the ratio of distances between the \textit{n}th and 1st nearest-neighbour distance on the undistorted square lattice.

We interpolate $V_{ij}(r,\sigma_i,\sigma_j)$ by introducing a new function $V_{ij}(r)$ defined such that $V_{ij}(r_H)=V_{ij}(r_H,1,1)$, $V_{ij}(r_L)=V_{ij}(r_L,-1,-1)$, and $V_{ij}(\overline{R})=V_{ij}(\overline{R},1,-1)=V_{ij}(\overline{R},-1,1)$, which yields
\begin{eqnarray}
\begin{split}
g_{ij}(r)=&V_{ij}(r-\overline{R}\eta_{ij})
-\left(r-\overline{R}\eta_{ij}\right)\left(\frac{\left(V_{ij}(r-\overline{R}\eta_{ij}+2\delta\eta_{ij} )-V_{ij}(r-\overline{R}\eta_{ij}-2\delta\eta_{ij})\right)}{4\delta\eta_{ij}}\right)\\
&+\frac{1}{2}\left(r-\overline{R}\eta_{ij}\right)^2\left(\frac{\left(V_{ij}(r-\overline{R}\eta_{ij}+2\delta\eta_{ij})-2V_{ij}(r-\overline{R}\eta_{ij})+V_{ij}(r-\overline{R}\eta_{ij}-2\delta\eta_{ij})\right)}{(2\delta\eta_{ij})^2}\right),\label{eq:V0}\\
\end{split}
\end{eqnarray}
\begin{eqnarray}
\begin{split}
h_{ij}(r)=&\left(\frac{\left(V_{ij}(r-\overline{R}\eta_{ij}+2\delta\eta_{ij} )-V_{ij}(r-\overline{R}\eta_{ij}-2\delta\eta_{ij})\right)}{4\delta\eta_{ij}}\right)\\
&-(r-\overline{R}\eta_{ij})\left(\frac{V_{ij}(r-\overline{R}\eta_{ij}+2\delta\eta_{ij})-2V_{ij}(r-\overline{R}\eta_{ij})+V_{ij}(r-\overline{R}\eta_{ij}-2\delta\eta_{ij})}{(2\delta\eta_{ij})^2}\right),\label{eq:h}
\end{split}
\end{eqnarray}
and
\begin{eqnarray}
\begin{split}
k_{ij}(r)&=\frac{V_{ij}(r-\overline{R}\eta_{ij}+2\delta\eta_{ij})-2V_{ij}(r-\overline{R}\eta_{ij})+V_{ij}(r-\overline{R}\eta_{ij}-2\delta\eta_{ij})}{(2\delta\eta_{ij})^2}.\label{eq:k}
\end{split}
\end{eqnarray}
Noting that 
\begin{equation}
V\left( r-\overline{R}\eta_{ij}+2\delta\eta_{ij} \right) = \sum_{n=0}^\infty V^{(n)}\left( r-\overline{R}\eta_{ij} \right)\frac{(2\delta\eta_{ij})^n}{n!},
\end{equation}
where
\begin{equation}
V^{(n)}(x)\equiv\left.\left(\frac{\partial^n V(r)}{\partial r^n} \right)\right|_{r=x},
\end{equation}
we find, with no further approximation, that
\begin{eqnarray}
g_{ij}(r)&=&V_{ij}(r-\overline{R}\eta_{ij})
-\left(r-\overline{R}\eta_{ij}\right)
\sum_{n=0}^\infty \frac{(2\delta\eta_{ij})^{2n}}{(2n+1)!} V_{ij}^{(2n+1)}\left( r-\overline{R}\eta_{ij} \right)
+\left(r-\overline{R}\eta_{ij}\right)^2 
\sum_{n=1}^\infty \frac{(2\delta\eta_{ij})^{2(n-1)}}{(2n)!} V_{ij}^{(2n)}\left( r-\overline{R}\eta_{ij} \right)
\\
&=&V_{ij}(r-\overline{R}\eta_{ij})
-\left(r-\overline{R}\eta_{ij}\right) V_{ij}^{(1)}\left( r-\overline{R}\eta_{ij} \right)
+ \frac{\left(r-\overline{R}\eta_{ij}\right)^2}{2} V_{ij}^{(2)}\left( r-\overline{R}\eta_{ij} \right) 
\notag \\&&
-\left(r-\overline{R}\eta_{ij}\right)
\frac{2(\delta\eta_{ij})^{2}}{3} V_{ij}^{(3)}\left( r-\overline{R}\eta_{ij} \right)
+ \left(r-\overline{R}\eta_{ij}\right)^2\frac{(\delta\eta_{ij})^2}{6} V_{ij}^{(4)}\left( r-\overline{R}\eta_{ij} \right) 
+\dots
\\
h_{ij}(r)
&=&\sum_{n=0}^\infty \frac{(2\delta\eta_{ij})^{2n}}{(2n+1)!} V_{ij}^{(2n+1)}\left( r-\overline{R}\eta_{ij} \right)
-2(r-\overline{R}\eta_{ij}) \sum_{n=1}^\infty \frac{(2\delta\eta_{ij})^{2(n-1)}}{(2n)!} V_{ij}^{(2n)}\left( r-\overline{R}\eta_{ij} \right),
\\
&=&
V_{ij}^{(1)}\left( r-\overline{R}\eta_{ij} \right)
- ( r-\overline{R}\eta_{ij} ) V_{ij}^{(2)}\left( r-\overline{R}\eta_{ij} \right) 
+ \frac{2(\delta\eta_{ij})^{2}}{3} V_{ij}^{(3)}\left( r-\overline{R}\eta_{ij} \right)
- ( r-\overline{R}\eta_{ij} ) \frac{(\delta\eta_{ij})^2}{3} V_{ij}^{(4)}\left( r-\overline{R}\eta_{ij} \right) +\dots \notag\\&&
\\
k_{ij}(r)&=& 2\sum_{n=1}^\infty \frac{(2\delta\eta_{ij})^{2(n-1)}}{(2n)!} V_{ij}^{(2n)}\left( r-\overline{R}\eta_{ij} \right)\\
&=& V_{ij}^{(2)}\left( r-\overline{R}\eta_{ij} \right) + \frac{(\delta\eta_{ij})^2}{3} V_{ij}^{(4)}\left( r-\overline{R}\eta_{ij} \right) 
+\dots
\end{eqnarray}
To reach Eqs. 4 and 5 of main text we set $f_{ij}(r)=g_{ij}(r)+h_{ij}(r)(r-\overline{R}\eta_{ij})$.

\section{ADDITIONAL RESULTS}

\begin{figure*}
	\includegraphics[scale=0.3]{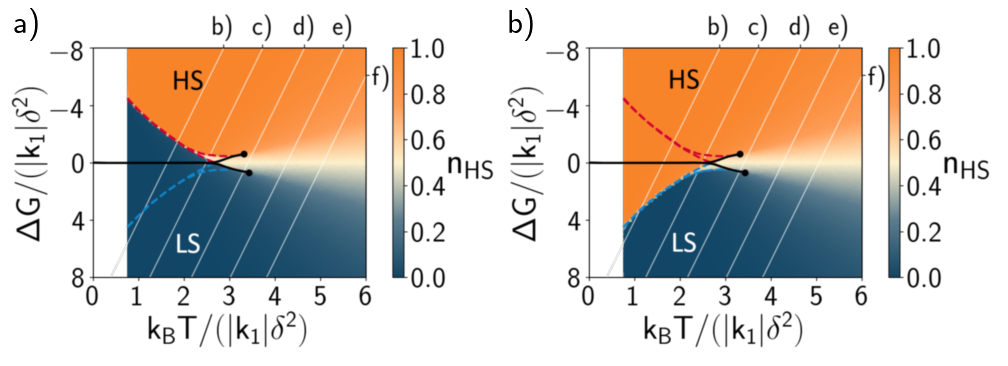} 
	\centering
	\caption{
		The HS fraction, $n_{HS}$, calculated on (a) cooling and (b) heating for the  nearest neighbour square lattice model with $k_1>0$. Lines and dots have the same meanings as in Fig. \ref{k1cv}a, where we show the full phase diagram.
	}
	\label{k1heatcool} 
\end{figure*}

In Figs. \ref{k1}a, \ref{OneStep}a, 5a, and 6a we show phase diagrams with lines indicating the limits of metastability on heating and cooling  at  fixed $\Delta H$. The calculations which these lines are based on are shown in (Fig. \ref{k1heatcool}, \ref{OneStepheatcool}, \ref{TwoStepNeelheatcool} and \ref{TwoStepStripeheatcool}).

\begin{figure*}
	\includegraphics[scale=0.3]{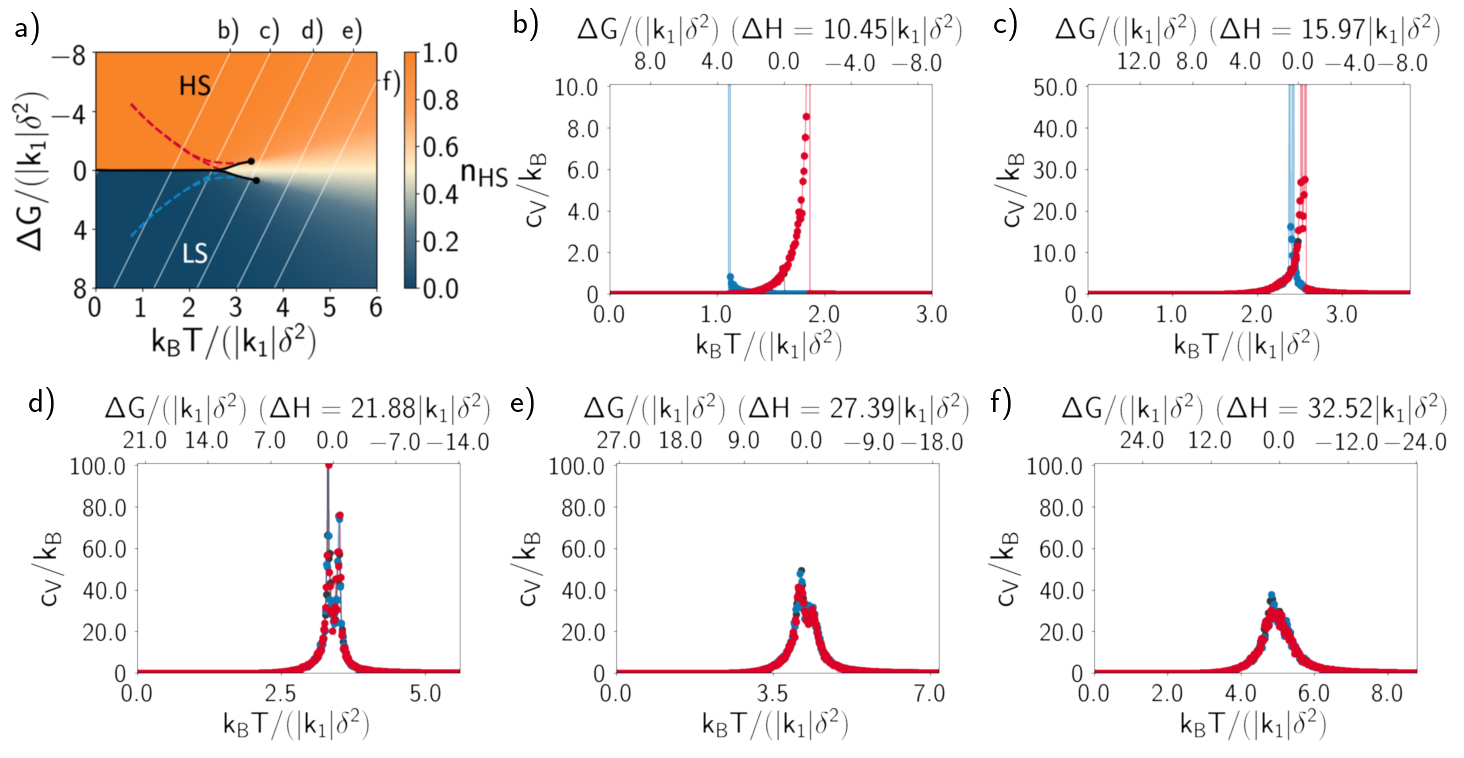}
	\centering
	\caption{(a) The phase diagram for nearest-neighbour interactions, $k_1>0$, reprinted from Fig. 4a for convenience.
		The  colours  indicate the equilibrium values for the fraction of high-spin metal centres, $n_{HS}$, calculated via parallel tempering. We find a (black) line of first order transitions that bifurcates at a triple point and ends in two critical points (black dots). There is no spontaneously broken symmetry or long-range order in the region between the two first order lines where $n_{HS}\simeq1/2$.
		The blue  (red) dashed line marks the limit of metastability for on the HS  (resp. LS) phases on  cooling (resp. heating), see Fig. \ref{k1heatcool}.
		Individual materials have fixed $\Delta H$ (not fixed $\Delta G$); the white lines are lines of constant $\Delta H$ and their labels correspond to panels (b-f) where the heat capacity, $c_V$, is plotted along these lines (see Fig. 4 for the corresponding  high spin fractions).
		In these plots the blue, red and black lines represent the cooling, heating and equilibrium  values respectively.
	}
	\label{k1cv} 
\end{figure*}

\begin{figure*}
	\includegraphics[scale=0.3]{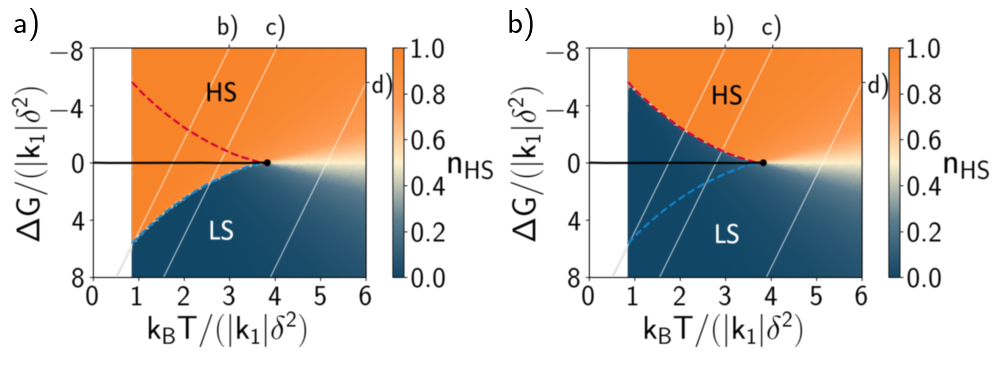} 
	\centering
	\caption{The HS fraction, $n_{HS}$, calculated on (a) cooling and (b) heating for the next nearest neighbour square lattice model with $k_2=0.1k_1>0$. Lines and dots have the same meanings as in Fig. \ref{k1cv}a. See Fig. \ref{OneStepcv}a for the full phase diagram.
	}
	\label{OneStepheatcool} 
\end{figure*}

\begin{figure*}
	\includegraphics[scale=0.28]{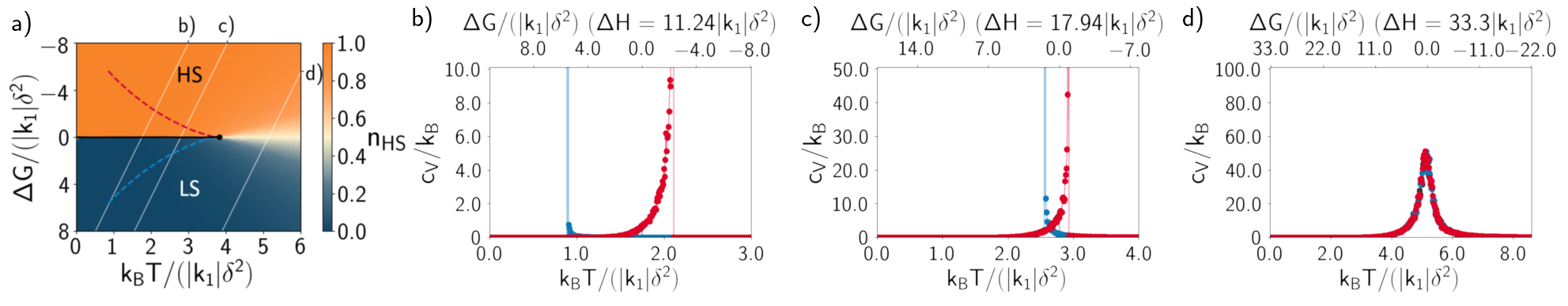} 
	\centering
	\caption{(a) Phase diagram for the next nearest neighbour square lattice model with $k_2=0.1k_1>0$ reprinted from Fig. \ref{OneStep}a for convenience. Lines and dots have the same meanings as in Fig. \ref{k1cv}a.
		(b-d) The heat capacity, $c_V$, (see Fig. \ref{OneStep} for the corresponding HS fractions). Blue, red and black lines show data for the cooling, heating and thermal equilibrium predictions, respectively.  
	}
	\label{OneStepcv} 
\end{figure*}

\begin{figure*}
	\includegraphics[scale=0.3]{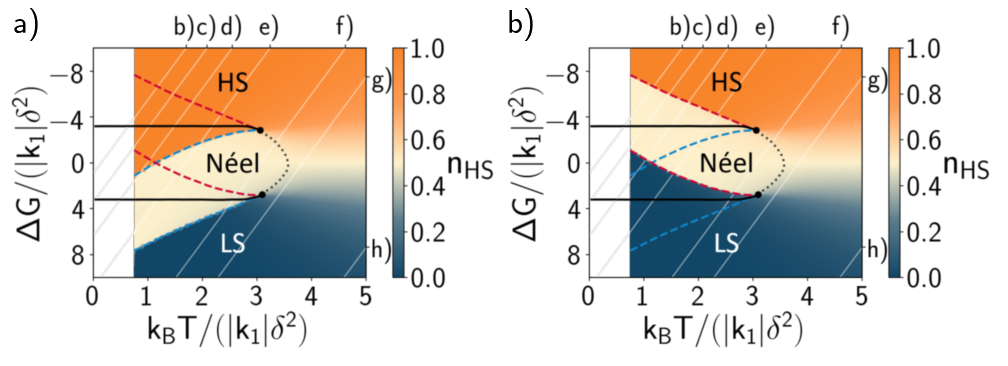} 
	\centering
	\caption{
		The HS fraction, $n_{HS}$, calculated on (a) cooling and (b) heating for the next nearest neighbour square lattice model  with $k_1>0$ and $k_2=-0.2{k_1}$, appropriate for the 1$n$24  and 1$n$02 families. Lines and dots have the same meanings as in Fig. \ref{k1cv}a. See Fig. \ref{TwoStepNeelcv}a for the full phase diagram.
	}
	\label{TwoStepNeelheatcool} 
\end{figure*}

\begin{figure*}
	\includegraphics[scale=0.25]{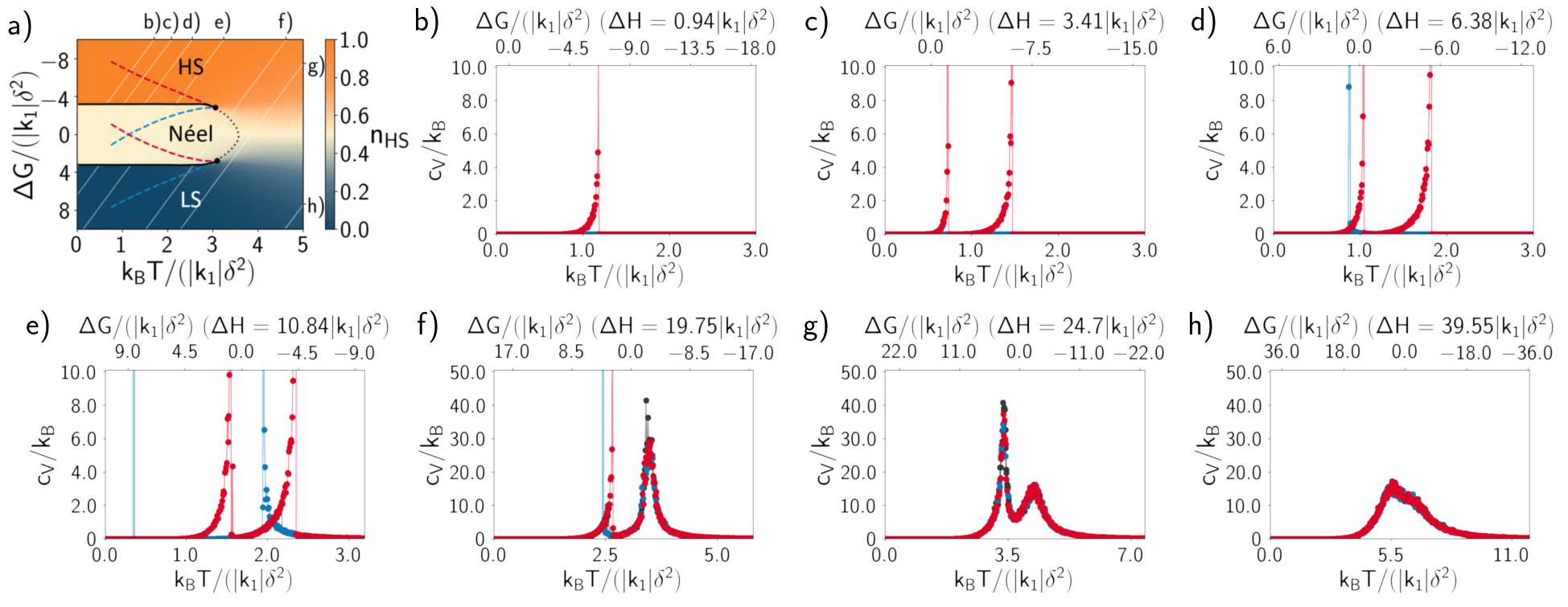} 
	\centering
	\caption{
		(a) Phase diagram for the next nearest neighbour square lattice model with $k_1>0$ and  $k_2=-0.2{k_1}$, appropriate for the 1$n$24  and 1$n$02 families, reprinted from Fig. 5a for convenience. Lines and dots have the same meanings as in Fig. \ref{k1cv}a.
		(b-h) The heat capacity, $c_V$, (see Fig. 5 for the corresponding HS fractions). Blue, red and black lines show data for the cooling, heating and thermal equilibrium predictions, respectively.  
	}
	\label{TwoStepNeelcv} 
\end{figure*}

\begin{figure*}
	\includegraphics[scale=0.3]{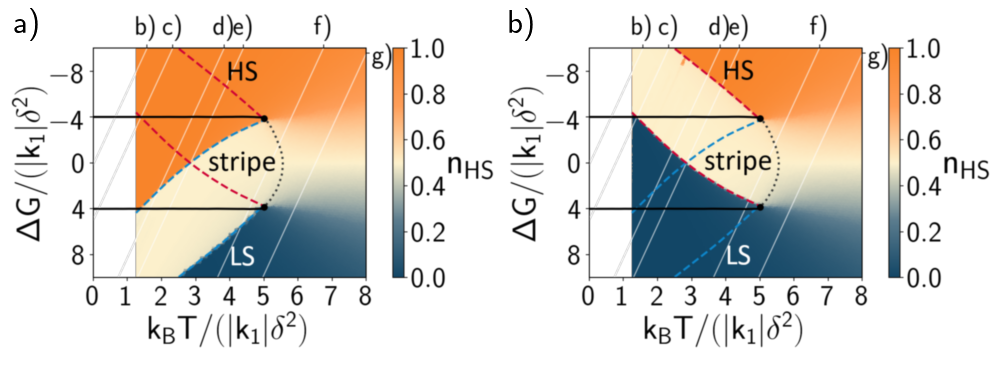} 
	\centering
	\caption{
		The HS fraction, $n_{HS}$, calculated on (a) cooling and (b) heating for the next nearest neighbour square lattice model  with $k_1<0$ and $k_2=1.2 |k_1|$, appropriate for the 1$n$14 family. Lines and dots have the same meanings as in Fig. \ref{k1cv}a. See Fig. \ref{TwoStepStripecv}a for the full phase diagram.
	}
	\label{TwoStepStripeheatcool} 
\end{figure*}

\begin{figure*}
	\includegraphics[scale=0.25]{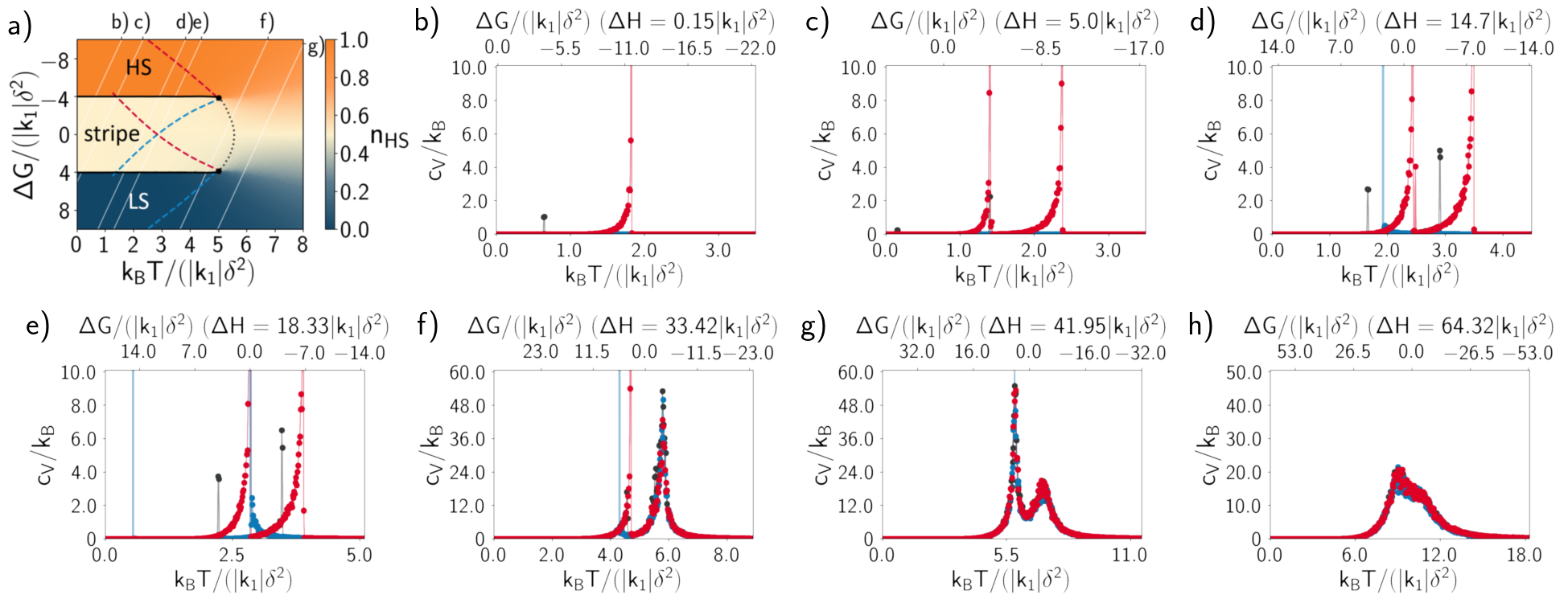} 
	\centering
	\caption{
		(a) Phase diagram for the next nearest neighbour square lattice model with $k_1<0$ and $k_2=1.2 |k_1|$, appropriate for the 1$n$14 family, reprinted from Fig. 6a for convenience. Lines and dots have the same meanings as in Fig. \ref{k1cv}a.
		(b-h) The heat capacity, $c_V$, (see Fig. 6 for the corresponding HS fractions). Blue, red and black lines show data for the cooling, heating and thermal equilibrium predictions, respectively.   
	}
	\label{TwoStepStripecv} 
\end{figure*}

\begin{figure*}
	\includegraphics[scale=0.25]{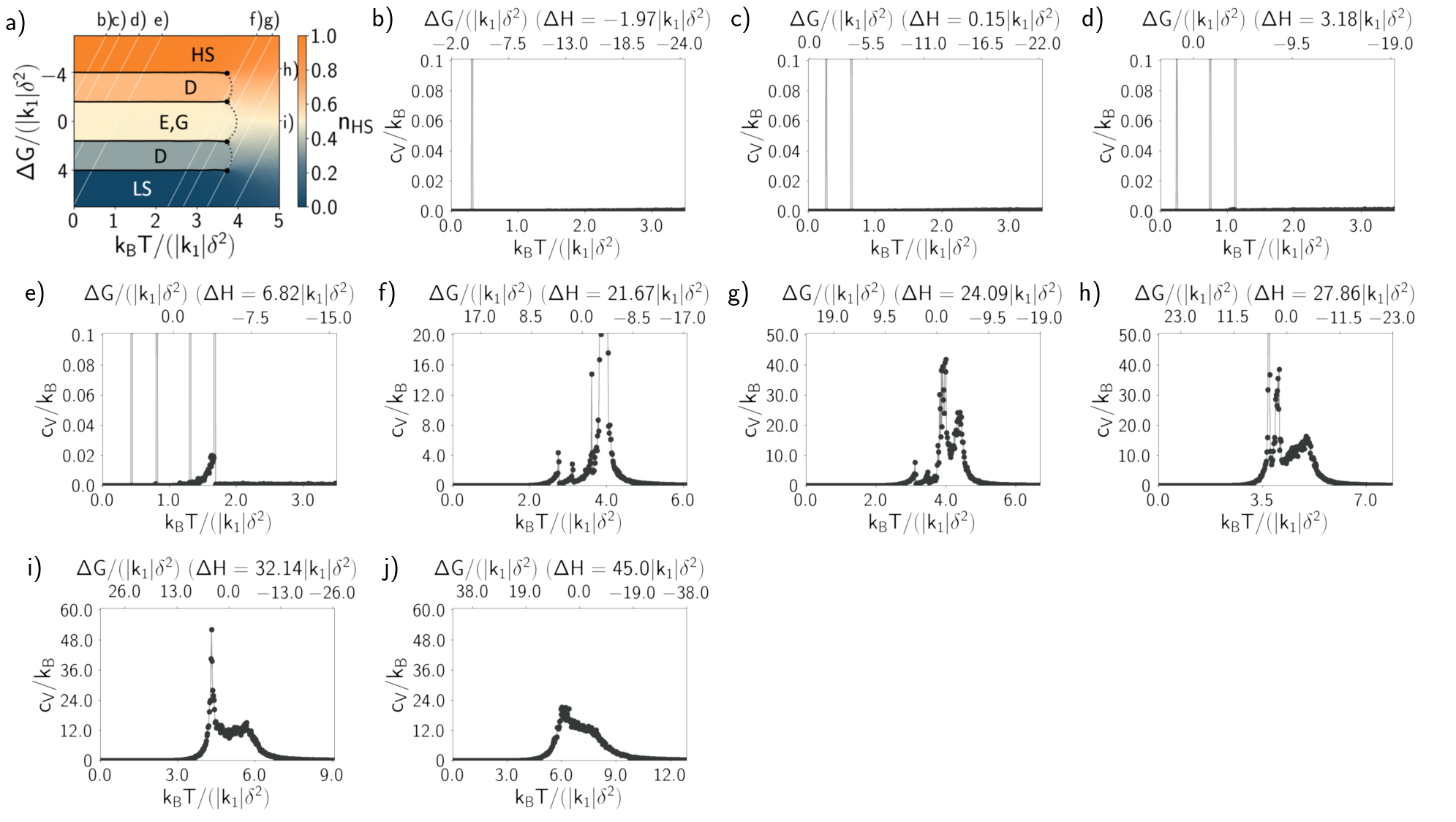} 
	\centering
	\caption{
		(a) Phase diagram for the third nearest neighbour square lattice model with $k_1>0$, $k_2=-0.9k_1$ and $k_3=0.5k_1$, appropriate for the 1$n$24  and 1$n$02 families, reprinted from Fig. 7a for convenience. Lines and dots have the same meanings as in Fig. \ref{k1cv}a.
		(b-j) The heat capacity, $c_V$, (see Fig. 7 for the corresponding HS fractions).  For simplicity  only  the parallel tempering results are shown. 
	}
	\label{FourStepBcv} 
\end{figure*}

\begin{figure*}
	\includegraphics[scale=0.25]{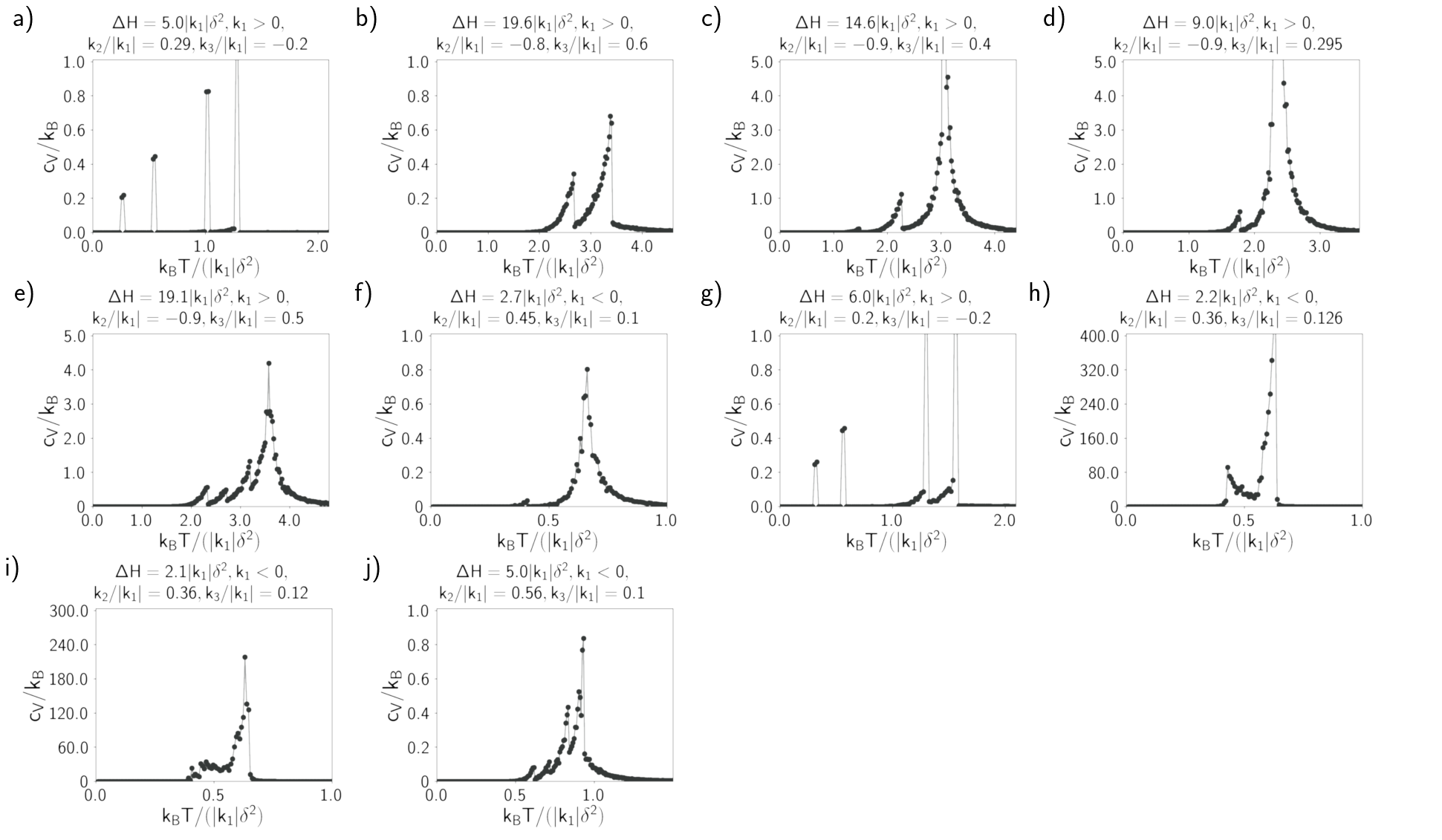} 
	\centering
	\caption{
		Temperature dependence of the heat capacity in the third nearest neighbour model with the same parameters for which $n_{HS}$ is plotted in Fig. 8. The parameters are reasonable for (a) 1$n$14 family, (b-e) 1$n$24  and 1$n$02 families  and (a-j)  supramolecular lattices. For simplicity we only show the parallel tempering results. 
	}
	\label{k3Transitionscv} 
\end{figure*}

In Figs. \ref{k1}b-f, \ref{OneStep}b-d, 5b-h, 6b-h, 7b-j, 8a-j,  and 10b-f of the main text we report the fraction of high spins as temperature varies The corresponding heat capacities, which provide a more sensitive signature of phase transitions and crossovers in SCO materials, are shown in Fig. \ref{k1cv}b-f, \ref{OneStepcv}b-d, \ref{TwoStepNeelcv}b-h, \ref{TwoStepStripecv}b-h, \ref{FourStepBcv}b-j, \ref{k3Transitionscv}a-j,  and \ref{EightStepscv}b-f respectively.

Lastly, in the main text we show the phase diagram for the next nearest neighbour 1$n$14 family lattice model with $k_1<1$, $k_2=1.2 |k_1|$ (Fig. 8,  see also Fig. \ref{TwoStepStripecv}). In Fig. \ref{Opt} we show phase diagram for the nearest neighbour model with $k_1<1$, $k_2=0.6 |k_1|$ demonstrating the importance of the relative contributions of the magnitudes of $k_1$ and $k_2$ on the thermodynamics.

\begin{figure*}
	\includegraphics[scale=0.26]{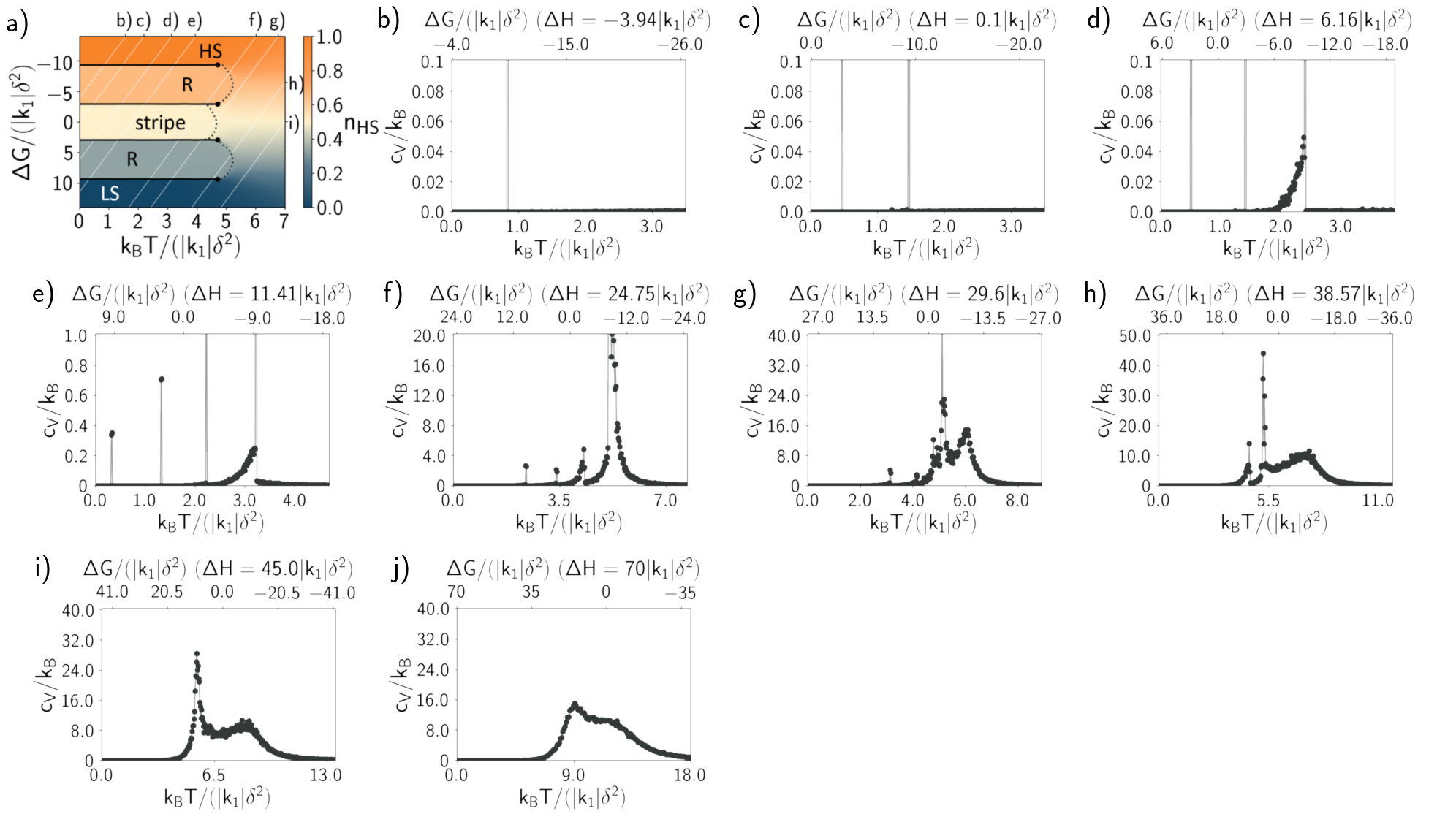} 
	\centering
	\caption{
		(a) Phase diagram for the fifth nearest neighbour square lattice model with $k_1<0$, $k_2=1.2 |k_1|$,  $k_3=-0.5 |k_1|$, $k_4=0$, and $k_5=0.2 |k_1|$, appropriate for the  1$n$14 family, reprinted from Fig. \ref{FourStepA}a for convenience. Lines and dots have the same meanings as in Fig. \ref{k1cv}a.
		(b-j) The heat capacity, $c_V$, (see Fig. \ref{FourStepA} for the corresponding HS fractions).  For simplicity  only  the parallel tempering results are shown.
	}
	\label{FourStepAcv} 
\end{figure*}

\begin{figure*}
	\includegraphics[scale=0.32]{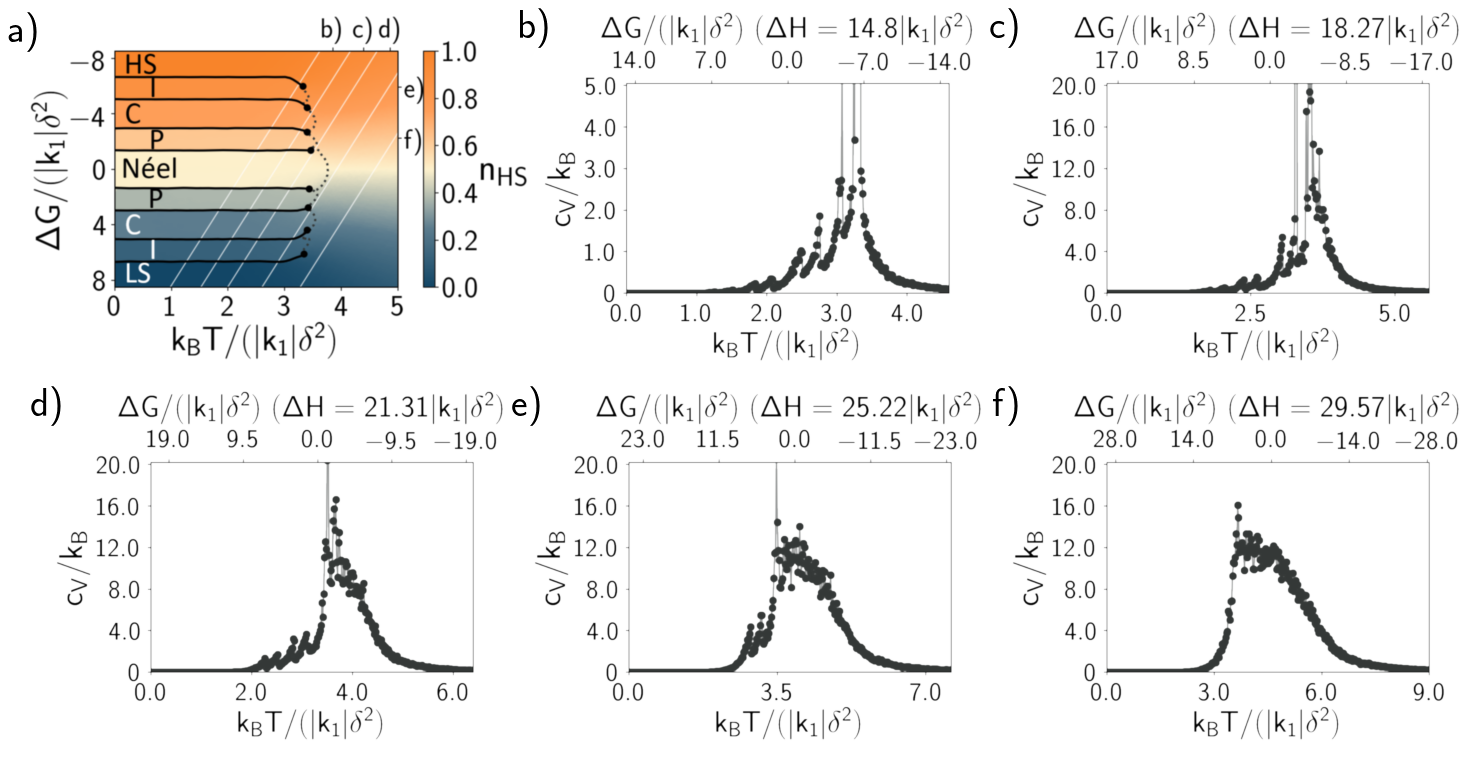} 
	\centering
	\caption{		
		(a) Typical slice of the finite temperature phase diagram all interactions up to fifth nearest neighbours. Lines and dots have the same meanings as in Fig. \ref{k1cv}a.
		(b-f) The heat capacity, $c_V$ (See Fig. 10 for the corresponding fraction of high spins). For simplicity only the parallel tempering predictions are shown.
	}
	\label{EightStepscv} 
\end{figure*}

\begin{figure*}
	\includegraphics[scale=0.28]{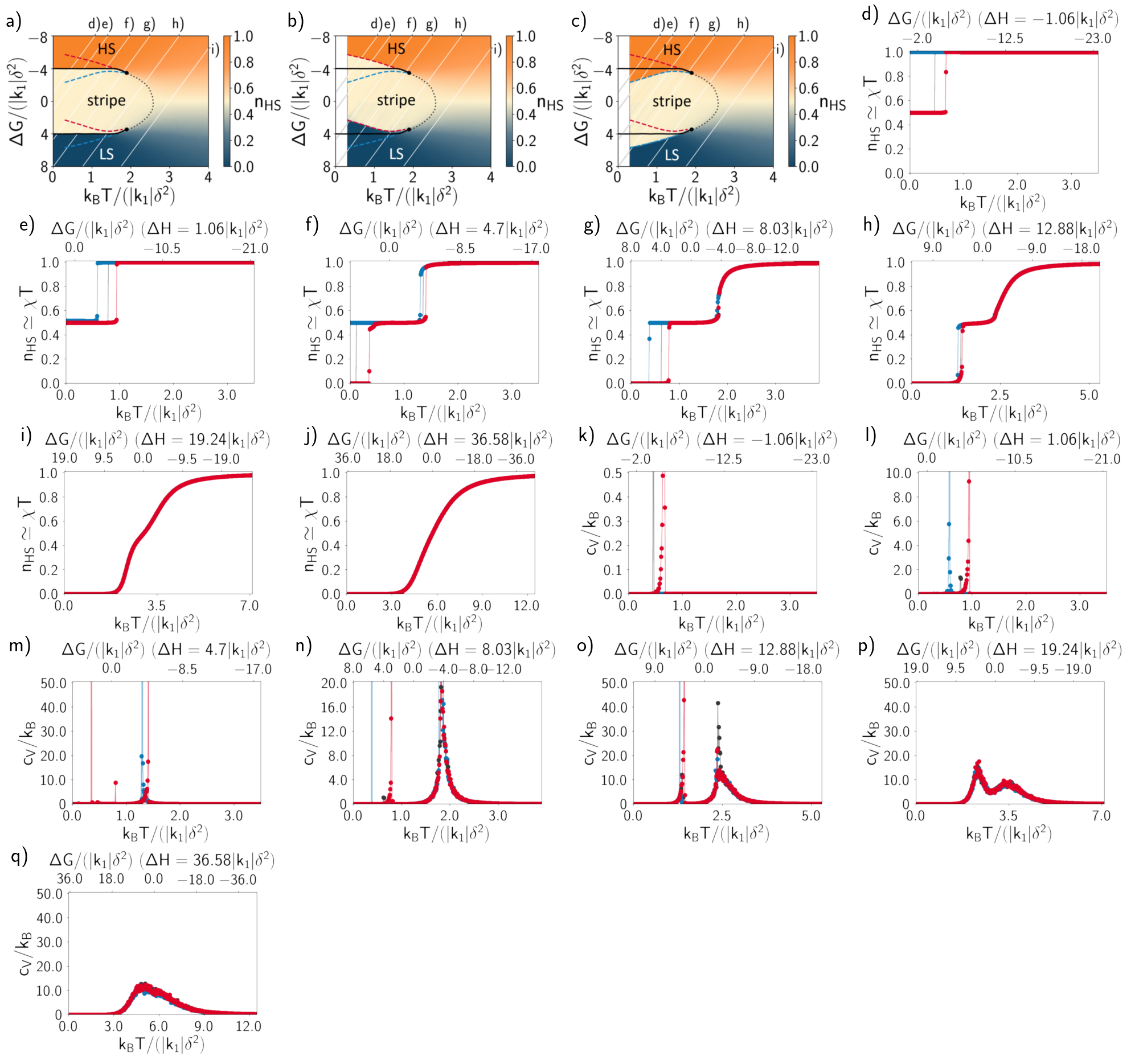} 
	\centering
	\caption{
		(a) Phase diagrams for the next nearest neighbour model with $k_1<0$, $k_2>0$ (here we take $k_2=0.6 |k_1|$,  other parameters give  similar results). Lines and dots have the same meanings as in Fig. \ref{k1cv}a. The HS fractions calculated on (b) heating and (c) cooling with fixed $\Delta H$ are used calculate the limits of stability. Note that in contrast to the results for $k_1>0$ and $k_2>0$ the relative magnitudes of $k_1$ and $k_2$ has importance consequences for the observed behaviours  (see, e.g., Figs. 6 and \ref{TwoStepStripecv}). 
		(d-j) The fraction of high spins, $n_{HS}$ and (k-q)  the corresponding heat capacities. 
		Blue, red and black lines show data for the cooling, heating and thermal equilibrium predictions, respectively. 
	}
	\label{Opt} 
\end{figure*}

\subsection{NEAREST-NEIGHBOUR INTERACTIONS}

\begin{figure*}
	\includegraphics[scale=0.3]{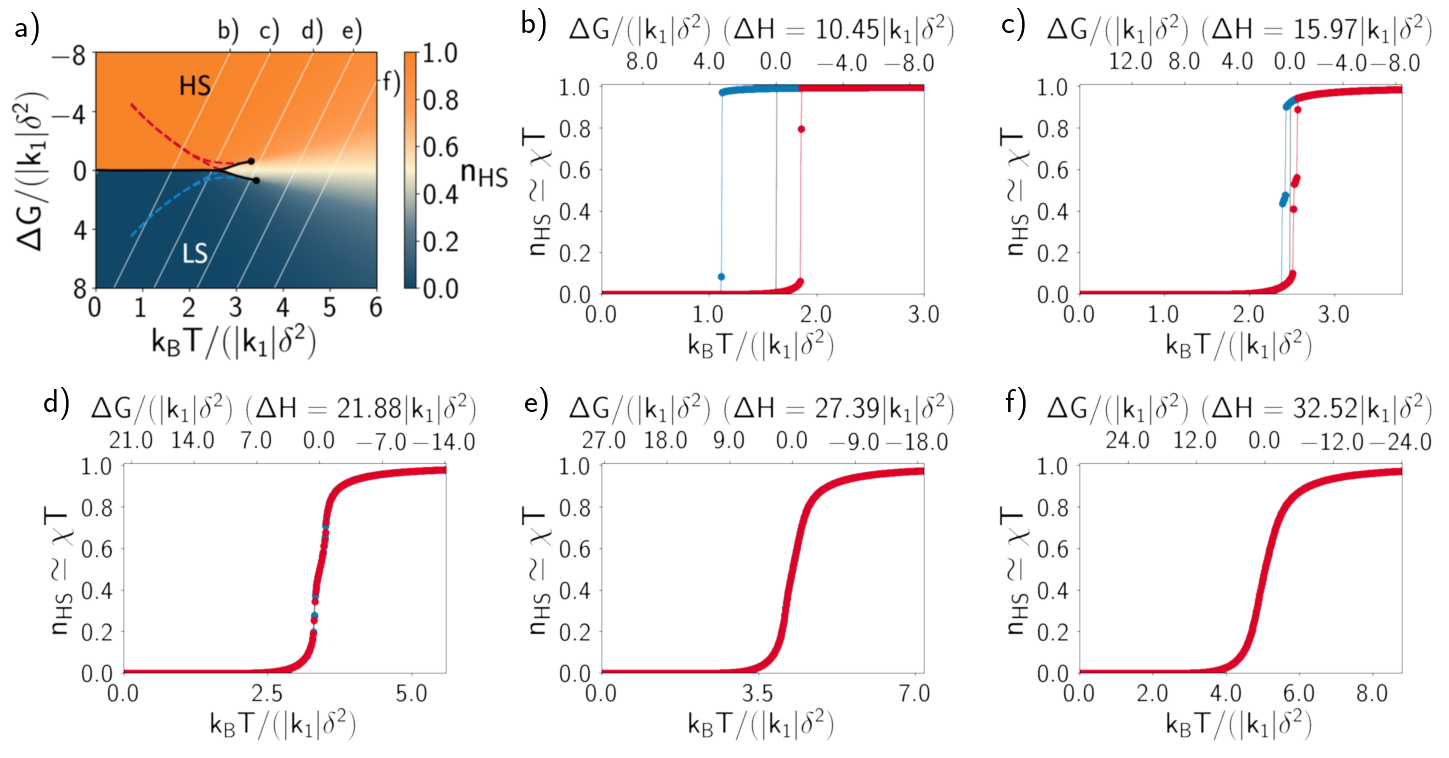} 
	\centering
	\caption{(a) The phase diagram for nearest-neighbour interactions, $k_1>0$. 
		Colours in the phase diagram indicate the fraction of high-spin $M$ sites, $n_{HS}\sim \chi T$ where $\chi$ is the susceptibility, calculated via parallel tempering. The (black) line of first order transitions  bifurcates at a triple point and ends in two critical points (black dots). There is no spontaneously broken symmetry or long-range order in the region between the two first order lines where $n_{HS}\simeq1/2$.
		The blue  (red) dashed line marks the limit of metastability for  the HS  (resp. LS) phases on  cooling (resp. heating), cf. Fig. S1, and show  the width of the hysteresis. Individual materials have fixed $\Delta H$,  white lines correspond to panels (b-f), where the fraction of high spins is plotted  (see Fig. S2 for the corresponding heat capacities).
		In these plots the blue, red and black lines represent the cooling, heating and parallel tempering values respectively.
	}
	\label{k1} 
\end{figure*}

In this section, we consider only the  nearest-neighbour elastic interaction, $k_1$. That is, we set $k_2=k_3=k_4=k_5=0$. The stability of the lattice requires  $k_1>0$.\cite{footnote} The  phase diagram of this model is shown in Fig. \ref{k1}a. As individual materials  have constant $\Delta H$, rather than constant $\Delta G$, we mark lines of constant $\Delta H$ on the phase diagram and report the thermodynamic properties at selected values of $\Delta H$ (Figs. \ref{k1}b-f).

When the single molecule contribution to the free energy is much greater than the contribution of the cooperative elastic interactions, $\Delta H \gg{k_1} \delta^2$,  we observe a gradual crossover (Figs. \ref{k1}f and S2f). This thermodynamic behaviour is commonly reported in experiments on weakly cooperative materials.\cite{gutlich}

In the opposite regime,   $\Delta H \ll{k_1} \delta^2$, we see a sharp one-step transition with hysteresis (Figs. \ref{k1}b and S2b). This indicates a first order phase transition, as is commonly observed in bistable SCO materials.\cite{gutlich} Note that in our calculations the first order spin-state transition is not accompanied by a crystallographic phase transition, which is excluded by the symmetric breathing mode approximation.

In the intermediate regime, $\Delta H/( |k_1| \delta^2)\sim 15-30$, we see two-steps. Depending on the magnitude of $\Delta H/( |k_1| \delta^2)$ these steps can occur as two first-order transitions with hysteresis (Figs. \ref{k1}c and S2c), one first order transition and one crossover (Figs. \ref{k1}d and S2d), or two crossovers (Figs. \ref{k1}e and S2e). In between these regimes are critical points where the transitions become continuous. 

The two-step behaviour is a result of the competition between the long-range strain and the elastic interactions. The elastic interactions favour an antiferroelastic phase with N\'eel ordering (Fig. 1a), whereas the long-range strain prefers all metal centres to be in the same spin-state. At $T=0$ when $\Delta H=\Delta G > 0$ the LS phase is realized and for $\Delta H=\Delta G < 0$ the HS phase is energetically favourable. However, for $T=0$ and $\Delta H=0$ the N\'eel ordered state is degenerate with the HS and LS states  (Figs. \ref{k1}a and S1a). At sufficiently high temperatures thermal fluctuations stabilize short-range N\'eel correlations for $\Delta G \approx 0$,  Figs. \ref{k1}a and S2a. This cannot stabilise a true, long-range ordered, N\'eel phase, but does result in a observed two-step transition, Figs. \ref{k1}d and S2d. 
The same qualitative features of this phase diagram have been also been captured by Chenyshov \textit{et al.} \cite{IsingChern} using a Landau theory.  

However, with only nearest neighbour interactions present, the N\'eel phase is the only antiferroelastic phase present. Furthermore, the N\'eel phase is observed only in an extremely narrow temperature range. Experimentally,  many different antiferroelastic phases have been found and these phases can stable over relatively broad temperature ranges. This is consistent with the idea that longer range elastic interactions are vitally important for multistep transitions.

\subsection{STRICTLY POSITIVE NEXT NEAREST NEIGHBOUR INTERACTIONS (ANY FAMILY)}

\begin{figure}
	\includegraphics[width=0.95\linewidth]{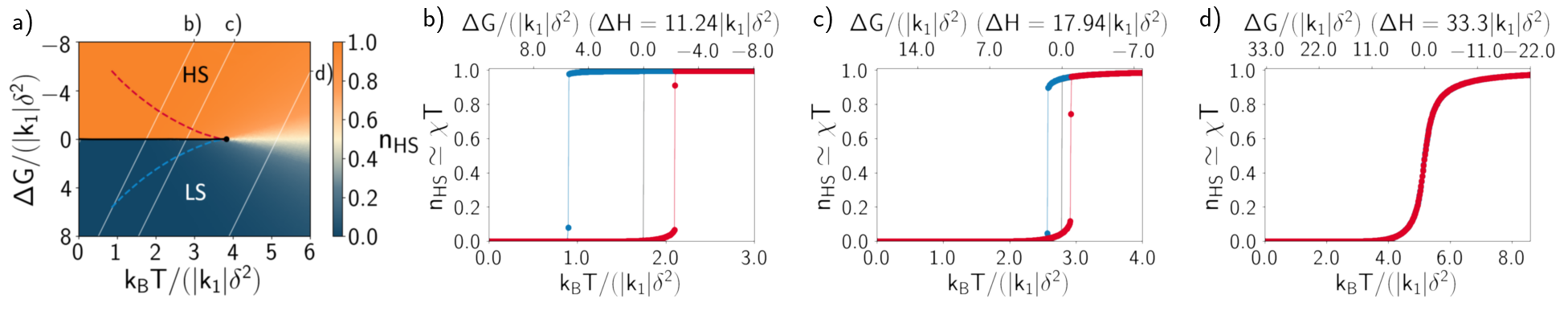} 
	\centering
	\caption{(a) Typical phase diagram for the next nearest neighbour square lattice model with $k_1$ and $k_2>0$ (here $k_2=0.1k_1$; see also Fig. S3). (b-d) The fraction of high spins, $n_{HS}$ (see Fig. S4 for the corresponding heat capacities). Symbols have the same meanings as in Fig. \ref{k1}. 
	}
	\label{OneStep} 
\end{figure}

$k_1>0$ and $k_2>0$ is possible for any of the lattices shown in Fig. 2 provided that the minima of both interactions are roughly commensurate with a square lattice.
At $T=0$ 
the LS and HS states are separated by a first order phase transition at $\Delta G=0$, Figs. 4a and \ref{OneStep}a. For $T>0$ this phase transition remains at $\Delta G=0$ until it reaches a critical point, where the transition is continuous. Thus, the frustration has entirely suppressed the N\'eel order found at finite temperatures for $k_2=0$ (compare Figs. \ref{k1} and \ref{OneStep}).

Considering lines of constant $\Delta H$, which represent individual materials, we see three distinct thermodynamic behaviours: When the elastic interactions are strong compared to the single molecule physics, $\Delta H\ll |k_1| \delta^2$, the transitions are sharp and first-order, the width of the hysteresis is  greater the smaller $\Delta H/ |k_1| \delta^2$ is (Figs. \ref{OneStep}b,c and S4b,c). If the elastic interactions are weak, $\Delta H\gg |k_1|\delta^2$, we find a crossover (Figs. \ref{OneStep}d and S4d). These two regimes are separated by a continuous (or second order) phase transition at the critical point. 

Consistent with this prediction, single step transitions are common and observed in all of the families of materials discussed here.

\begin{figure*}
	\includegraphics[scale=0.275]{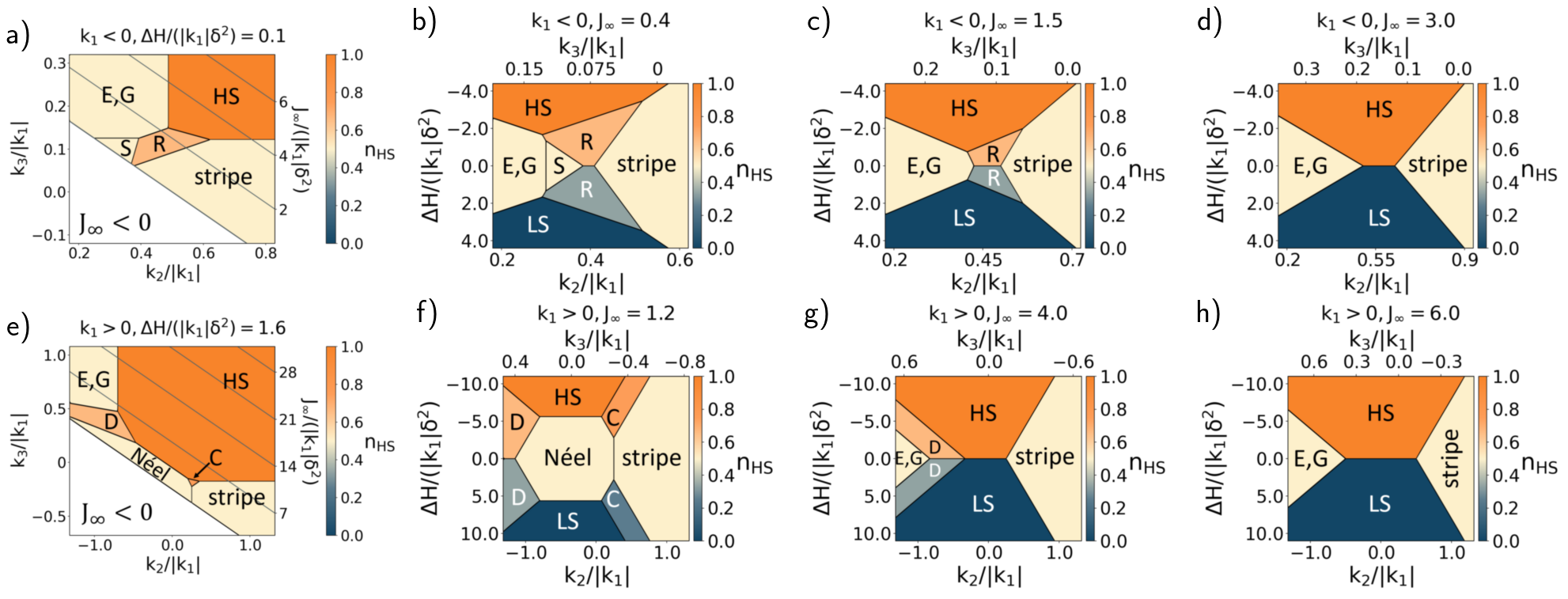} 
	\centering
	\caption{Slices of the zero-temperature phase diagram with up to third nearest-neighbour elastic interactions $k_1$, $k_2$ and $k_3$ for (a,e) constant $\Delta H$ and (b-d,f-h) constant $J_\infty$. Grey lines in (a,e) indicate lines of constant $J_\infty$. In addition to the phases that are energetically preferred by individual elastic interactions (HS, LS, N\'eel, stripe, C, E and G), frustration  introduces additional phases into the zero-temperature phase diagram (D, R and S). E, G indicates that the E and G phases are degenerate. 
	} 
	\label{k3ZeroTemp} 
\end{figure*}

\begin{figure*}
	\includegraphics[scale=0.35]{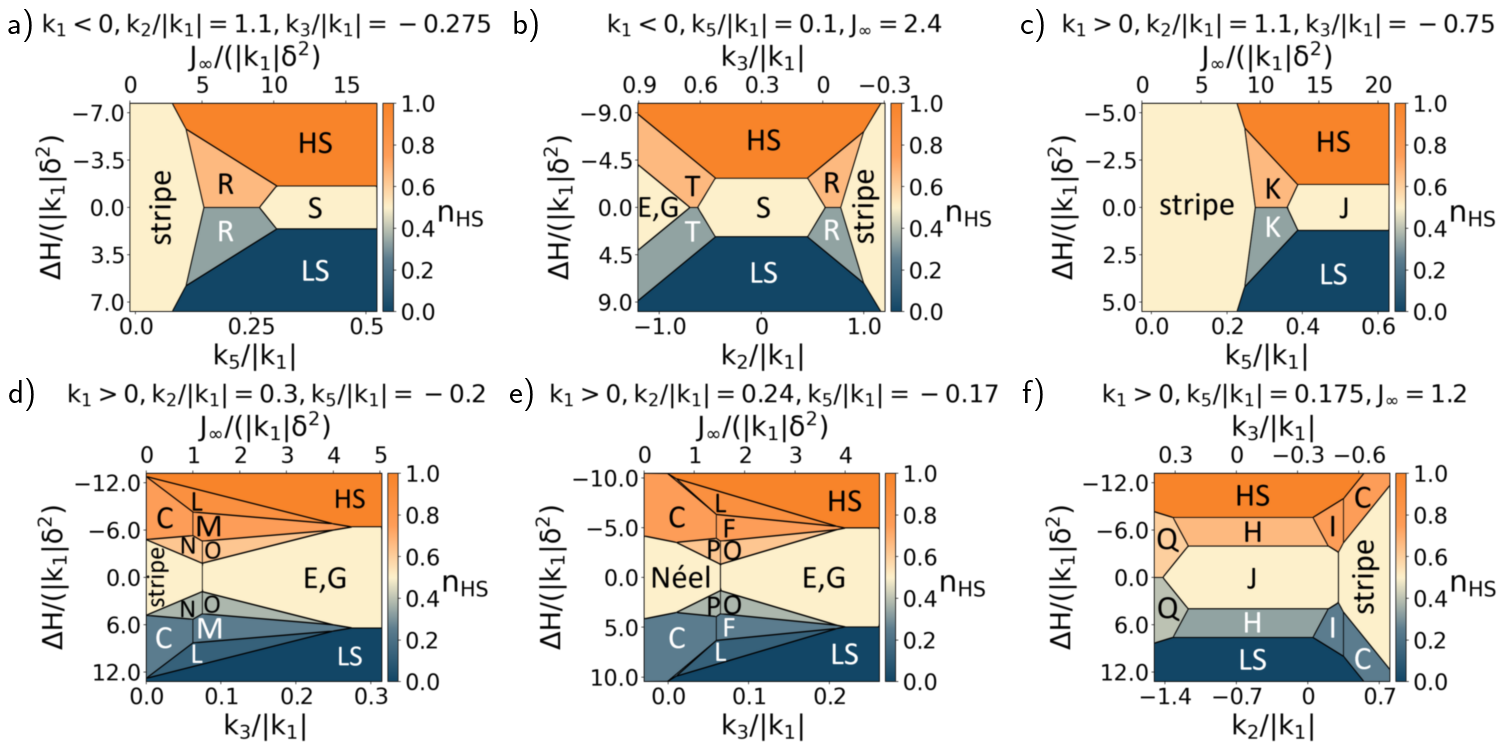} 
	\centering
	\caption{Selected slices of the zero temperature phase diagram with up to fifth nearest-neighbour interactions showing thirty-six distinct phases. In general increasing the number of nearest-neighbour interactions increases the complexity of the phase diagram.
	}
	\label{k5ZeroTemperature} 
\end{figure*}

\begin{figure*}
	\includegraphics[scale=0.25]{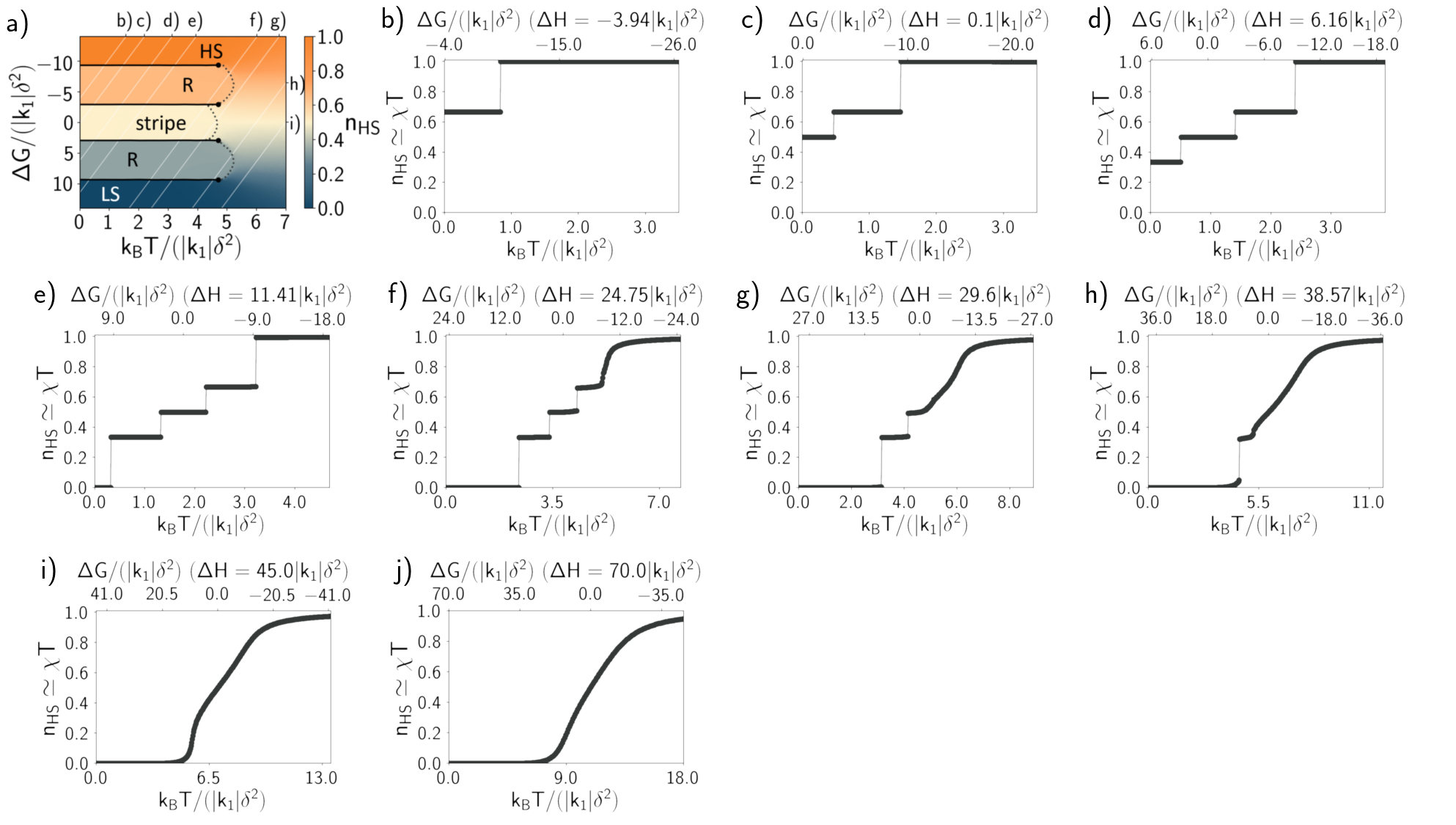} 
	\centering
	\caption{	(a) Typical phase diagram for the 1$n$14 family   with $k_1<0$, $k_2>0$, $k_3<0$, $k_4=0$ and $k_5>0$ (here $k_2=1.2 |k_1|$,  $k_3=-0.5 |k_1|$ and $k_5=0.2 |k_1|$).  The R phase consists of alternating stripes of width 2 and width 1 (Fig. 1). Lines and dots have the same meanings as in Fig. \ref{k1cv}a.
		(b-i) The fraction of high spins, $n_{HS}$ (see Fig. \ref{FourStepAcv} for the corresponding heat capacities). Only  the   parallel tempering Monte Carlo predictions are reported. 
	}
	\label{FourStepA} 
\end{figure*}


\begin{thebibliography}{99}
\bibitem{footnote}
The lattice is stable to shear distortions because the symmetric breathing mode approximation is equivalent to assuming an infinite spring constant for shear modes.
\end{thebibliography}
\end{document}